\documentclass[seceq]{ptptex}

\usepackage{graphicx}
\usepackage{wrapft}

\preprintnumber[3cm]{%<-- [..]: optional width of preprint # column.
NAOJ-Th-Ap 2003,No.4\\PTPTeX ver.0.8\\ February, 2003}
\title{%        %You can use \\ for explicit line-break
  Comparison of the Oscillatory Behaviors of a Gravitating Nambu-Goto
  String and a Test String 
}

\author{%       %Use \scshape  for the family name
  Kouji {\sc Nakamura}\footnote{E-mail: kouchan@th.nao.ac.jp} 
}

\inst{%         %Affiliation, neglected when [addenda] or [errata]
  Division of Theoretical Astrophysics,
  National Astronomical Observatory, \\
  Tokyo, Mitaka 181-8588, Japan
}

\recdate{February 14, 2003}%            %Editorial Office will fill in this.

\abst{%         %this abstract is neglected when [addenda] or [errata]
  Comparison of the oscillatory behavior of a gravitating infinite
  Nambu-Goto string and a test string is investigated using the
  general relativistic gauge invariant perturbation technique
  with two infinitesimal parameters on a flat spacetime
  background.
  Due to the existence of the pp-wave exact solution, we see that
  the conclusion that the dynamical degree of freedom of an infinite
  Nambu-Goto string is completely determined by that of gravitational
  waves, which was reached in our previous works [K.~Nakamura,
  A.~Ishibashi and H.~Ishihara, Phys.~Rev.\ D{\bf 62} (2002), 101502(R);
  K.~Nakamura and H.~Ishihara, Phys.~Rev.\ D{\bf 63} (2001), 127501.],
  do not contradict to the dynamics of a test string.
  We also briefly discuss the implication of this result.
}

\begin{document}

\maketitle

%%%%%%%%%%%%%%%%%%%%%%%%%%%%%%%%%%%%%%%%%%%%%%%%%%%%%%%%%%%%%%%%%%%%%%
\section{Introduction}
\label{sec:introduction}
%%%%%%%%%%%%%%%%%%%%%%%%%%%%%%%%%%%%%%%%%%%%%%%%%%%%%%%%%%%%%%%%%%%%%%

Nambu-Goto membranes are one of the simplest models of the extended
objects that appear in various physical contexts.
Particularly, the gravitational fields of extended objects provide 
interesting and important topics, including the dynamics of
topological and/or non-topological defects (domain walls and cosmic
strings\cite{Vilenkin_Shellard,Vilenkin_Everett}) and the
gravitational waves emitted from them, the simplest type of brane
worlds (vacuum brane\cite{RS}).

%*********************************************\\

It is well known that if the self-gravity of membranes is ignored,
which is called ``the test membrane case,'' the equation of motion
derived from the Nambu-Goto action admits oscillatory solutions,
and membranes oscillate freely.
However, recent investigations using exactly soluble
models\cite{Kodama,kouchan-cyl,Ishibashi-Tanaka} have revealed that
the oscillatory behavior of gravitating Nambu-Goto walls differs from
that of test walls. In particular, gravitating walls cannot oscillate
freely. 
The main points of these works are essentially separated into two parts: 
First, they showed that {\it the dynamical degree of freedom of
the perturbative wall oscillations is completely determined by that of
the gravitational waves.} 
Second, considering the gravitational wave scattering by a
gravitating wall, they showed that {\it there is no resonance
  pole, except for damping modes.}

%*********************************************\\

Such analysis was later extended to the case of the perturbative
oscillations of an infinite Nambu-Goto
string\cite{kouchan-string-1,kouchan-string-2}. 
The results of this extension are summarized as follows.
{\it The dynamical degree of freedom of the string oscillations
  is completely determined by that of gravitational waves} (as in the
wall cases), and {\it an infinite string can oscillate continuously,
  but such oscillations simply represent the propagation of
  gravitational waves along the string}. 
The gravitational waves along the string are just the pp-wave exact
solution to the Einstein equation, called cosmic string traveling
waves. 
Note that the existence of the pp-wave solution is closely related to
the specific symmetry of the spacetime. 
Therefore, it is natural to conjecture that the existence of the
continuous oscillations of an infinite string is due to the special
symmetry of the model and that the traveling wave solutions do not
exist in more generic situations. 
Of course, this needs to be proven and it is merely speculation here.

%*********************************************\\

In any case, from the results of perturbation analyses around exact
solutions, it is reasonable to regard that the dynamical degree of
freedom of string oscillations is that of gravitational waves.
This implies that (A) {\it the oscillatory behavior of gravitating
  membranes might differ from those of test membranes even if their
  energy densities are very small}, because the background spacetimes
of these models are exact solutions to the Einstein equation.
However, if we regard the Nambu-Goto string as an idealization
of GUT scale cosmic strings, their deficit angle is roughly
approximated to be $\sim 10^{-6}$. 
Then, the following conjecture also seems natural.
(B) {\it If the energy densities of membranes are very small, it
  should be possible to describe their gravitational fields as
  perturbations with a Minkowski spacetime background and describe
  their dynamics as those of test membranes.}
However, the statements (A) and (B) seem to contradict each other.

%*********************************************\\

The aim of this article is to compare the oscillatory behavior of a
gravitating string and that of a test string.  
We concentrate on the dynamics of an infinite Nambu-Goto string, using
general relativistic perturbation theory on a Minkowski background
spacetime.
We develop the gauge invariant perturbation theory with two
infinitesimal parameters on the Minkowski spacetime background.
We denote one of the perturbation parameters by $\epsilon$, which
corresponds to the string oscillation amplitude, and the other by 
$\lambda$, which corresponds to the string energy density.
To determine the motion of the string, we solve the Einstein equation
order by order. 
Physically, the perturbation in $\epsilon$ describes the oscillations
of a string without its gravitational field. 
This corresponds to the oscillations of a test string.
The perturbations in $\lambda$ describes the gravitational field of an
infinite static string, without oscillations.
The oscillations of a gravitating string, the emission of gravitational
waves due to these oscillations, and its radiation reaction are
described by the simultaneous perturbations in $\epsilon$ and $\lambda$.

%*********************************************\\

In this article, we gives the perturbative analyses to $O(\epsilon)$,
$O(\lambda)$, and $O(\epsilon\lambda)$ to compare with the analyses in
our previous works\cite{kouchan-string-1,kouchan-string-2}. 
The previous works are based on the first order perturbation with
respect to the string oscillation amplitude.
Therefore, $O(\epsilon\lambda)$ perturbations are the lowest order
perturbations that are appropriate to compare with the oscillatory
behavior of a gravitating string.
We show that the $O(\epsilon\lambda)$ Einstein equations
are almost same as those of the perturbative equations in our
previous work. 
Further, we find that due to the existence of the above-mentioned
pp-wave exact solution, the oscillations of a test infinite string are
not contrary to the dynamics of a gravitating infinite string, at least
to first order in the oscillation amplitude of the string.

%*********************************************\\

The organization of this article is as follows.
In the next section (\S\ref{sec:self-gravitating-NG-string}), we
briefly comment on the energy-momentum tensor for the regularized
Nambu-Goto string in order to study the dynamics of a gravitating
Nambu-Goto string.
To consider a gravitating string, we should consider a regularized
string rather than an infinitesimally thin string.
In \S\ref{sec:two-parameter-perturbation-in-GR}, the gauge invariant
perturbation technique with two infinitesimal parameters in general
relativity is developed.
In \S\ref{sec:epsilon-order-lambda-order-solution}, we give the
solutions to the $O(\epsilon)$ and $O(\lambda)$ Einstein equations. 
The $O(\epsilon\lambda)$ Einstein equations are given in
\S\ref{sec:Perturbed-Einstein-Equations}.  
Comparison with the results of our previous
works\cite{kouchan-string-1,kouchan-string-2} is made in
\S\ref{sec:Comparison}.  
The final section (\S\ref{sec:summary-discussion}) is devoted to
a summary and discussion.

%*********************************************\\

Throughout this paper, we denote Newton's gravitational constant
by $G$ and we use units such that the light velocity $c$ is $1$.
Further, we use abstract index notation\cite{Wald-book}.

%*********************************************\\

%%%%%%%%%%%%%%%%%%%%%%%%%%%%%%%%%%%%%%%%%%%%%%%%%%%%%%%%%%%%%%%%%%%%%%
\section{Self-gravitating regularized Nambu-Goto string}
\label{sec:self-gravitating-NG-string}
%%%%%%%%%%%%%%%%%%%%%%%%%%%%%%%%%%%%%%%%%%%%%%%%%%%%%%%%%%%%%%%%%%%%%%

%*********************************************\\

To consider the dynamics of gravitating extended objects of
co-dimension two (or larger), we must treat a delicate problem in
general relativity.  
It is well known that there is no simple prescription of an arbitrary
line source where a metric becomes singular\cite{Geroch-Traschen}. 
Although it seems plausible that the string dynamics are well
approximated by the dynamics of conical singularities, it has been
shown that the world sheet of conical singularities must be 
totally geodesic\cite{Unruh}. This is called Israel's Paradox. 
This implies that it is impossible for a generic Nambu-Goto string
can be idealized in terms of conical singularities.

%*********************************************\\

One of the simplest procedures to avoid this delicate problem is to
introduce the string thickness\cite{kouchan-loop-initial}. 
As discussed in our previous
papers\cite{kouchan-string-1,kouchan-string-2}, we first consider a
thick Nambu-Goto string, in which the singularity is regularized. 
Using this regularized string, we consider the perturbative dynamics of
the thick string by solving the Einstein equations.
Next, as in our previous
works\cite{kouchan-string-1,kouchan-string-2}, we consider the
situation of a thin string, in which the string thickness is much
smaller than the curvature scale of the bending string world sheet.

%*********************************************\\

The energy-momentum tensor of a ``thick string'' is obtained as the
extension of that for an infinitesimally thin string (see Appendix
\ref{sec:Appendix-thick-2}).
Let us consider a four-dimensional spacetime $({\cal M},g_{ab})$
including a ``thick string.''
The energy-momentum tensor for an infinitesimally thin string has
support only on its world sheet $\Sigma_{2}$, which is a
two-dimensional hypersurface in ${\cal M}$.
When we consider the spacetime including an infinitesimally thin
string, we assume that the region in the neighborhood of $\Sigma_{2}$
in the spacetime ${\cal M}$ (at least in the neighborhood of
$\Sigma_{2}$) can be foliated by two-dimensional surfaces.
The tangent space of the spacetime ${\cal M}$ in the neighborhood of
$\Sigma_{2}$ can be decomposed into two two-dimensional subspaces, as
seen in Appendix \ref{sec:Appendix-thick-1}. 
To introduce the thickness of the string, we consider a compact region
${\cal D}$ in the complement space of $\Sigma_{2}$, and we regard
the ``thick string world sheet'' to be ${\cal D}\times\Sigma_{2}$
(see Appendix \ref{sec:Appendix-thick-3}).

%*********************************************\\

In this article, we consider the energy-momentum tensor defined by 
\begin{eqnarray}
  \label{eq:NG-energy-momentum-tensor}
  T_{ab} := - \rho q_{ab}
\end{eqnarray}
as that for the ``thick string,'' where $\rho$ is the string energy
density and $q_{ab}$ is an extension of the intrinsic metric on
$\Sigma_{2}$. 
The string energy density $\rho$ and the intrinsic metric $q_{ab}$ are 
extended so that they have support on the ``thick string world sheet''
${\cal D}\times\Sigma_{2}$, as seen in Appendix 
\ref{sec:Appendix-thick-3}.
The metric $\gamma_{ab}$ on the complement space of the ``thick
string'' is defined by 
\begin{equation}
  \gamma_{ab} := g_{ab} - q_{ab}.
\end{equation}
The rank of both the metrics $q_{ab}$ and $\gamma_{ab}$ is two.
The string thickness is characterized by the support of the energy
density $\rho$.

%*********************************************\\

The divergence of this energy-momentum tensor, 
$\nabla_{b}T_{a}^{\;\;b} = 0$, is given by 
\begin{eqnarray}
  \label{eq:NG-continuity-eq}
  && q^{ab}\nabla_{b}\rho + \rho \gamma^{bc}\nabla_{b} q_{c}^{\;\;a} = 0, \\
  \label{eq:NG-eq-of-motion}
  && \rho q^{bc}\nabla_{b} q_{c}^{\;\;a} =: \rho K^{a} = 0,
\end{eqnarray}
where $K^{a}$ is the extrinsic curvature of the ``thick string
world sheet'' defined in Appendix \ref{sec:Appendix-thick}. 
The Equation (\ref{eq:NG-continuity-eq}) is the continuity
equation, which arises due to the introduction of the string
thickness, and Eq.~(\ref{eq:NG-eq-of-motion}) is identical to with the
equation of motion derived from the Nambu-Goto action. 
Hence, to study the dynamics of a gravitating Nambu-Goto string, we
can concentrate only on the Einstein equation with the energy-momentum
tensor (\ref{eq:NG-energy-momentum-tensor}).  
After solving the Einstein equation, we consider the thin string
situation, if necessary.

%*********************************************\\

%%%%%%%%%%%%%%%%%%%%%%%%%%%%%%%%%%%%%%%%%%%%%%%%%%%%%%%%%%%%%%%%%%%%%%
\section{General relativistic two-parameter perturbation}
\label{sec:two-parameter-perturbation-in-GR}
%%%%%%%%%%%%%%%%%%%%%%%%%%%%%%%%%%%%%%%%%%%%%%%%%%%%%%%%%%%%%%%%%%%%%%

%*********************************************\\

Here, we develop the perturbation theory with two perturbation
parameters in general relativity.
In this article, we consider the perturbative oscillations of an
infinite Nambu-Goto string using the Einstein equation with the above
energy-momentum tensor (\ref{eq:NG-energy-momentum-tensor}).
The background for the perturbation considered here is the Minkowski
spacetime, and we have two infinitesimal parameters for the
perturbation. 
One corresponds to the string oscillation amplitude, denoted by
$\epsilon$, and the other corresponds to the string energy density, 
denoted by $\lambda$.

%*********************************************\\

As mentioned in the Introduction (\S\ref{sec:introduction}), we give
the analyses for $O(\epsilon)$, $O(\lambda)$, and
$O(\epsilon\lambda)$ perturbations.
As the second order perturbation in addition to $O(\epsilon\lambda)$,
we may consider $O(\lambda^{2})$ and $O(\epsilon^{2})$ perturbations. 
$O(\lambda^{2})$ perturbations describe the static gravitational field
of an infinite string and $O(\epsilon^{2})$ perturbations describe the
test string oscillations of second order with respect to the string
oscillation amplitude. 
Clearly, these perturbations have nothing to do with gravitational
waves.
Further, since our analyses are order by order, $O(\lambda^{2})$ and
$O(\epsilon^{2})$ perturbations do not affect the $O(\epsilon\lambda)$
results. 
These are independent of each other.
For these reasons, we do not consider  $O(\lambda^{2})$ and
$O(\epsilon^{2})$ perturbations here.

%*********************************************\\

%%%%%%%%%%%%%%%%%%%%%%%%%%%%%%%%%%%%%%%%%%%%%%%%%%%%%%%%%%%%%%%%%%%%%%
\subsection{Background spacetime and its tangent space}

%*********************************************\\

Let us denote the background Minkowski spacetime by 
$({\cal M}_{0},\eta_{ab})$.
When we consider the perturbative oscillations of the string, it
is convenient to decompose the entire background spacetime
into two submanifolds.
One of the submanifolds contains the world sheet of an infinite
straight string.
We denote this submanifold by $({\cal M}_{1},\bar{q}_{ab})$.
The other submanifold is the complement space of ${\cal M}_{1}$.
We denote this submanifold by $({\cal M}_{2},\bar{\gamma}_{ab})$.
The entire background spacetime is given by the product of these
manifolds, i.e., ${\cal M}_{0} = {\cal M}_{1}\times{\cal M}_{2}$.

%*********************************************\\

The metric $\bar{q}_{ab}$ on ${\cal M}_{1}$ is the intrinsic metric of
the straight test string. The tensor 
$\bar{q}_{a}^{\;\;b} = \bar{q}_{ac}\eta^{cb}$
projects the tensor fields on the tangent space of the background
${\cal M}_{0}$ into those of ${\cal M}_{1}$.
We also introduce the indices $i,j,\cdots$ for the components of the
tangent space of ${\cal M}_{1}$, so that
\begin{equation}
  \label{eq:latine-indices-def}
  \bar{q}_{ab} = \bar{q}_{ij}(d\sigma^{i})_{a}(d\sigma^{j})_{b} 
  = - (dt)_{a}(dt)_{b} + (dz)_{a}(dz)_{b}.
\end{equation}
The metric $\bar{\gamma}_{ab}$ on ${\cal M}_{2}$ is defined by
$\bar{\gamma}_{ab} := \eta_{ab}-\bar{q}_{ab}$. 
We also introduce the indices $\alpha,\beta,\cdots$ for the components
of the tangent space of the complement space, which is normal to
the background string world sheet, so that 
\begin{equation}
  \label{eq:greek-indices-def}
  \bar{\gamma}_{ab} 
  = \bar{\gamma}_{\alpha\beta}(d\xi^{\alpha})_{a}(d\xi^{\beta})_{b} 
  = (dx)_{a}(dx)_{b} + (dy)_{a}(dy)_{b}.
\end{equation}
We also apply the Einstein convention for the summation over the
indices $i,j,\cdots$ and $\alpha,\beta,\cdots$.
This is different from the abstract index notation developed in
the textbook by Wald\cite{Wald-book}.

%*********************************************\\

Henceforth, we denote the metric $\bar{q}_{ab}$ ($\bar{q}_{ij}$) on
${\cal M}_{1}$ (and its components) by $q_{ab}$ ($q_{ij}$), for
simplicity. 
Further, we denote the metrics $\bar{\gamma}_{ab}$ on ${\cal M}_{2}$
and its components by simply $\gamma_{ab}$ and $\gamma_{\alpha\beta}$,
respectively.
We also denote the covariant derivative associated with the background
metrics $\eta_{ab}$, $q_{ab}$ and $\gamma_{ab}$ by $\nabla_{a}$,
${\cal D}_{a}$ and $D_{a}$, respectively.
Though this notation was originally introduced for use on the
physical spacetime ${\cal M}$, we use it throughout this paper except
in \S\ref{sec:self-gravitating-NG-string} and Appendix
\ref{sec:Appendix-thick}.

%*********************************************\\

%%%%%%%%%%%%%%%%%%%%%%%%%%%%%%%%%%%%%%%%%%%%%%%%%%%%%%%%%%%%%%%%%%%%%%
\subsection{Perturbative variables}

%*********************************************\\

Now, we introduce perturbative variables for the $O(\epsilon)$,
$O(\lambda)$, and $O(\epsilon\lambda)$ perturbations.

%*********************************************\\

The spacetime metric $g_{ab}$ is expanded as 
\begin{eqnarray}
  \label{eq:metric-perturbation}
  g_{ab} = \eta_{ab} + \epsilon h_{ab} + \lambda l_{ab}
  + \epsilon\lambda k_{ab}.
\end{eqnarray}
We can calculate the perturbative curvatures of $O(\epsilon)$,
$O(\lambda)$ and $O(\epsilon\lambda)$ straightforwardly by starting
from this expansion.

%*********************************************\\

The physical meanings of the infinitesimal perturbation parameters
$\epsilon$ and $\lambda$ are clarified when we consider the
perturbations of the energy-momentum tensor.
We expand the energy-momentum tensor for an infinite Nambu-Goto string 
as follows:
\begin{eqnarray}
  \label{eq:energy-momentum-perturbation}
  T_{a}^{\;\;b} = - \lambda \rho q_{a}^{\;\;b} 
  - \epsilon\lambda \left( 
    \rho \delta q_{a}^{\;\;b}
    + 
    \delta\rho q_{a}^{\;\;b} 
  \right).
\end{eqnarray}

%*********************************************\\

Since we consider the perturbation with Minkowski spacetime
background, the energy density $\rho$ of the string is also a
perturbative variable.
To make explicit that ``$\rho$ is small'', we replace $\lambda\rho$ in
the energy-momentum tensor (\ref{eq:energy-momentum-perturbation}) by
$\rho$.
Hence, $\lambda$ is the infinitesimal perturbation parameter that
corresponds to the energy density of the string.

%*********************************************\\

The $O(\epsilon)$ perturbation $\delta q_{a}^{\;\;b}$ of the intrinsic
metric $q_{a}^{\;\;b}$ in Eq.~(\ref{eq:energy-momentum-perturbation})
is given by 
\begin{eqnarray}
  \label{eq:intrinsic-perturbation}
  \delta q_{a}^{\;\;b} = \gamma_{ac} q_{d}^{\;\;b} h^{cd} 
  - \gamma_{\alpha\beta} \eta^{db} (\zeta^{\alpha})_{d} q_{a}^{\;\;c}
  \nabla_{c}\xi_{1}^{\beta} 
  - \gamma_{\alpha\beta} (\zeta^{\alpha})_{a} q_{c}^{\;\;b}
  \nabla^{c} \xi_{1}^{\beta},
\end{eqnarray}
\begin{figure}[htbp]
%\begin{wrapfigure}{l}{0.55\textwidth}
%  \includegraphics[width=0.48\textwidth]{Progfig-1.eps}
  \begin{center}
    \includegraphics[width=0.7\textwidth]{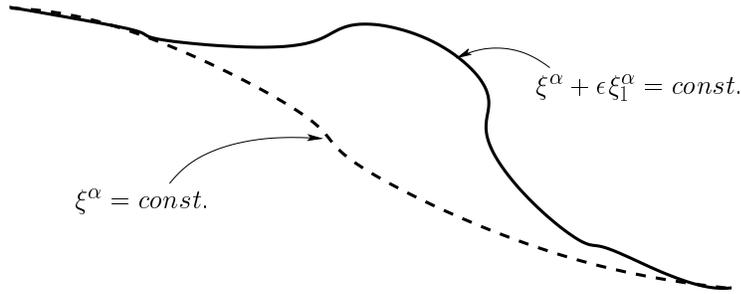}
    \caption{Deformation of the string world sheet. The dashed curve
      is the background string world sheet $\xi^{\alpha}=$const, and
    the solid curve is the perturbed string world sheet
    $\xi^{\alpha}+\epsilon\xi^{\alpha}_{1}=$const.}  
  \label{fig:JGRGfig3-1}
  \end{center}
%\end{wrapfigure}
\end{figure}
where
$(\zeta^{\alpha})_{a}:=\nabla_{a}\xi^{\alpha}=(d\xi^{\alpha})_{a}$
is the coordinate basis introduced in Eq.~(\ref{eq:greek-indices-def}).
The background string world sheet is the two-dimensional surface
on which all coordinate functions $\xi^{\alpha}$ are constant.
As depicted in Fig. \ref{fig:JGRGfig3-1}, the perturbed string world
sheet is also the two-dimensional surface on which
$\xi^{\alpha}+\epsilon\xi^{\alpha}_{1}$ is constant.
The second and third terms on the right-hand side of
Eq.~(\ref{eq:intrinsic-perturbation}) are induced by the perturbative
displacement $\xi_{1}^{\alpha}$.  
Thus, the parameter $\epsilon$ is associated with the string
oscillation amplitude. 
Since we regard the metric perturbation $h_{ab}$ as $O(\epsilon)$,
we should consider the perturbation $h_{ab}$ to be induced by the
perturbative displacement $\xi_{1}^{\alpha}$.

%*********************************************\\

Further, $\delta\rho$ in Eq.~(\ref{eq:energy-momentum-perturbation})
should be regarded as $O(\epsilon\lambda)$.
We should consider the continuity equation
(\ref{eq:NG-continuity-eq}) as being due to the introduction of the
string thickness. 
Equation (\ref{eq:NG-continuity-eq}) implies that the perturbative
deformation of the string induces a perturbative energy density. 
Since the energy density $\rho$ is the $O(\lambda)$ perturbative
variable itself, the first-order perturbation of the energy density 
should be regarded as $O(\epsilon\lambda)$ or $O(\lambda^{2})$ order.
Since $O(\lambda^{2})$ perturbations have nothing to do with the
dynamics of the string, we regard $\delta\rho$ in
Eq.~(\ref{eq:energy-momentum-perturbation}) to be $O(\epsilon\lambda)$.

%*********************************************\\

Starting from the perturbative expansions
(\ref{eq:metric-perturbation}) and 
(\ref{eq:energy-momentum-perturbation}), we calculate the perturbative
Einstein equations order by order. 
We also note that each order metric perturbation includes gauge
freedom that is irrelevant to the physical perturbations.  
The perturbative variables $\delta\rho$ and $\xi_{1}^{\alpha}$ also
include gauge freedom.
To study the oscillations of the string and compare with the result
in our previous work\cite{kouchan-string-1,kouchan-string-2}, we must
completely exclude such gauge freedom.
To do this, we develop the perturbation theory with two infinitesimal
parameters in a gauge invariant manner. 
To accomplish this, we first give that gauge transformation
rules of the perturbative variables from general point of view.
Then, we construct the gauge invariant variables for the perturbations
defined by
(\ref{eq:metric-perturbation})--(\ref{eq:intrinsic-perturbation}).

%*********************************************\\

%%%%%%%%%%%%%%%%%%%%%%%%%%%%%%%%%%%%%%%%%%%%%%%%%%%%%%%%%%%%%%%%%%%%%%
\subsection{Gauge transformations}

%*********************************************\\

Let us consider the background spacetime $({\cal M}_{0},\eta_{ab})$
and a physical spacetime $({\cal M},g_{ab})$, which we attempt to
describe as a perturbation of the background spacetime.
In relativistic perturbation theory, we are accustomed to the
expressions of the forms (\ref{eq:energy-momentum-perturbation})
and (\ref{eq:metric-perturbation}) that are the relations between a
tensor field (such as the metric) on the physical spacetime, the
background value of the same field, and its perturbation. 
Let us formally represent these expressions by 
\begin{equation}
  \label{eq:variable-symbolic-perturbation}
  Q(x) = Q_{0}(x) + \delta Q(x),
\end{equation}
symbolically. 
In this expression, we are implicitly assigning a correspondence
between points of the perturbed and the background spacetimes.
Moreover, the perturbed and unperturbed variables at the ``same''
point ``$x$'' are also defined implicitly. 
In the expression (\ref{eq:variable-symbolic-perturbation}), we
are implicitly considering the map 
${\cal M}_{0}\rightarrow{\cal M}$$:$
$p\in{\cal M}_{0}\mapsto q\in{\cal M}$.
This correspondence associated with the map 
${\cal M}_{0}\rightarrow{\cal M}$ is what is usually called a
gauge choice in the context of perturbation
theory\cite{J.M.Stewart-book}.
Clearly, this is more than the usual assignment of coordinate
labels to points of the single spacetime. 
Furthermore, the correspondence established by relations such as
(\ref{eq:variable-symbolic-perturbation}) is not unique by
itself, but, rather, (\ref{eq:variable-symbolic-perturbation})
involves the degree of freedom of the choice of the map
${\cal M}_{0}\rightarrow{\cal M}$ (i.e., the choice of the point
identification map ${\cal M}_{0}\rightarrow{\cal M}$). 
This is called gauge freedom.
Further, this freedom always exists in the perturbation of a theory
in which we impose general covariance.

%*********************************************\\

Following this understanding of gauge freedom, Bruni et
al\cite{Bruni-gr-qc-0207105} derived the gauge transformation rules up
to fourth order. 
Here, we give their $O(\epsilon)$, $O(\lambda)$, and
$O(\epsilon\lambda)$ results.
At these orders, we must consider three kinds of gauge
transformation rules for the perturbation of a physical variable
$Q$ on the physical manifold ${\cal M}$. 
We denote $O(\epsilon)$, $O(\lambda)$ and $O(\epsilon\lambda)$
order perturbations under the gauge 
${\cal X}:{\cal M}\rightarrow{\cal M}_{0}$ by 
${}^{{\cal X}}\delta^{(\epsilon)}Q$, 
${}^{{\cal X}}\delta^{(\lambda)}Q$ and 
${}^{{\cal X}}\delta^{(\epsilon\lambda)}Q$, respectively. 
We consider the gauge transformation 
$\Phi:= {\cal X}^{-1}\circ{\cal Y}: {\cal M}_{0}\rightarrow
{\cal M}_{0}$ (the gauge transformation from ${\cal X}$ to ${\cal Y}$). 
Under this gauge transformation, the perturbative variables are
transformed as follows: 
\begin{eqnarray}
  {}^{{\cal Y}}\delta^{(\epsilon)}Q - {}^{{\cal X}}\delta^{(\epsilon)}Q
  &=& {\pounds}_{{}^{(\epsilon)}\xi}Q_{0},
  \nonumber\\
  \label{eq:gauge-transformations}
  {}^{{\cal Y}}\delta^{(\lambda)}Q - {}^{{\cal X}}\delta^{(\lambda)}Q
  &=& {\pounds}_{{}^{(\lambda)}\xi}Q_{0},
  \\
  {}^{{\cal Y}}\delta^{(\epsilon\lambda)}Q - {}^{{\cal X}}\delta^{(\epsilon\lambda)}Q
  &=& 
  {\pounds}_{{}^{(\epsilon)}\xi} {}^{{\cal X}}\delta^{(\lambda)}Q 
  + {\pounds}_{{}^{(\lambda)}\xi} {}^{{\cal X}}\delta^{(\epsilon)}Q 
  \nonumber\\
  && \quad\quad
  + \left\{ {\pounds}_{{}^{(\epsilon\lambda)}\xi}
    + \frac{1}{2}
    {\pounds}_{{}^{(\epsilon)}\xi}{\pounds}_{{}^{(\lambda)}\xi} 
    + \frac{1}{2}
    {\pounds}_{{}^{(\lambda)}\xi}{\pounds}_{{}^{(\epsilon)}\xi} 
  \right\} Q_{0},
  \nonumber    
\end{eqnarray}
where ${}^{(\epsilon)}\xi$, ${}^{(\lambda)}\xi$ and 
${}^{(\epsilon\lambda)}\xi$ are the $O(\epsilon)$, $O(\lambda)$
and $O(\epsilon\lambda)$ generators of the gauge
transformation $\Phi$,
respectively. 
Gauge transformations of $O(\epsilon)$ and $O(\lambda)$ have
well-known forms.
Though the results derived by Bruni et al\cite{Bruni-gr-qc-0207105} include
superfluous parameters at $O(\epsilon\lambda)$, it is easy to check
that these parameters can be fixed through the replacement of the
$O(\epsilon\lambda)$ generator ${}^{(\epsilon\lambda)}\xi$ of
$\Phi$, and doing so, the transformation rules 
(\ref{eq:gauge-transformations}) are obtained\cite{kouchan-two-parameters}.
In terms of the passive coordinate transformation, the above gauge
transformation $\Phi$ is given by 
\begin{eqnarray}
  \tilde{x}^{\mu} &=& x^{\mu}(q)
  - \lambda {}^{(\lambda)}\xi^{\mu}
  - \epsilon {}^{(\epsilon)}\xi^{\mu}
  \nonumber\\
  && \quad
  + \epsilon\lambda \left\{ 
    - {}^{(\epsilon\lambda)}\xi^{\mu}
    + \frac{1}{2} {}^{(\lambda)}\xi^{\nu}
    \partial_{\nu}{}^{(\epsilon)}\xi^{\mu} 
    + \frac{1}{2} {}^{(\epsilon)}\xi^{\nu}
    \partial_{\nu}{}^{(\lambda)}\xi^{\mu} 
  \right\}.
\end{eqnarray}
Inspecting the transformation rules
(\ref{eq:gauge-transformations}), we develop a gauge invariant
perturbation theory on Minkowski spacetime.

%*********************************************\\

%%%%%%%%%%%%%%%%%%%%%%%%%%%%%%%%%%%%%%%%%%%%%%%%%%%%%%%%%%%%%%%%%%%%%%
\subsection{Gauge invariant variables}

%*********************************************\\

Inspecting the gauge transformation rules
(\ref{eq:gauge-transformations}), we define the gauge invariant
perturbative variables from the perturbed variables $h_{ab}$,
$l_{ab}$, $k_{ab}$, $\xi_{1}^{a}$ and $\delta\rho$.
To carry this out, we use the procedure to construct the gauge
invariant variables developed in a forthcoming
paper\cite{kouchan-two-parameters}. 
The $O(\lambda)$ string energy density $\rho$ is also
a perturbative variable.
Since the background value of the string energy density is
trivial\cite{J.M.Stewart-book}, $O(\lambda)$ string energy density
$\rho$ is gauge invariant itself.

%*********************************************\\

First, we consider the metric perturbations $h_{ab}$ and $l_{ab}$ of
$O(\epsilon)$ and $O(\lambda)$, respectively.
Under the gauge transformation $\Phi={\cal X}^{-1}\circ{\cal Y}$,
these metric perturbations are transformed as    
\begin{eqnarray}
  \label{eq:epsilon-metric-gauge-trans}
  {}^{{\cal Y}}h_{ab} - {}^{{\cal X}}h_{ab} &=& 2
  \nabla_{(a}{}^{(\epsilon)}\xi_{b)}, \\
  \label{eq:lambda-metric-gauge-trans}
  {}^{{\cal Y}}l_{ab} - {}^{{\cal X}}l_{ab} &=& 2
  \nabla_{(a}{}^{(\lambda)}\xi_{b)}.
\end{eqnarray}
The metric perturbations $h_{ab}$ and $l_{ab}$ are decomposed as
\begin{eqnarray}
  \label{eq:epsilon-metric-decomp}
  h_{ab} &=:& {\cal H}_{ab} + 2 \nabla_{(a}{}^{(\epsilon)}X_{b)}, \\
  \label{eq:lambda-metric-decomp}
  l_{ab} &=:& {\cal L}_{ab} + 2 \nabla_{(a}{}^{(\lambda)}X_{b)},
\end{eqnarray}
where the variables ${\cal H}_{ab}$, ${}^{(\epsilon)}X_{a}$, 
${\cal L}_{ab}$ and ${}^{(\lambda)}X_{a}$ are
transformed as 
\begin{eqnarray}
  && {}^{{\cal Y}}{\cal H}_{ab} - {}^{{\cal X}}{\cal H}_{ab} = 0,
  \quad
  {}^{{\cal Y}(\epsilon)}X_{a}-{}^{{\cal X}(\epsilon)}X_{a} =
  {}^{(\epsilon)}\xi_{a},
  \\
  && {}^{{\cal Y}}{\cal L}_{ab} - {}^{{\cal X}}{\cal L}_{ab} = 0,
  \quad
  {}^{{\cal Y}(\lambda)}X_{a}-{}^{{\cal X}(\lambda)}X_{a} =
  {}^{(\lambda)}\xi_{a} 
\end{eqnarray}
under the gauge transformation $\Phi={\cal X}^{-1}\circ{\cal Y}$.
Clearly, ${\cal H}_{ab}$ (${\cal L}_{ab}$) is the gauge invariant
part, and ${}^{(\epsilon)}X_{b}$ (${}^{(\lambda)}X_{b}$) is the gauge
variant part of the metric perturbation $h_{ab}$ ($l_{ab}$).
These decompositions are accomplished by carrying out the mode
expansion, as shown in \S\ref{sec:lambda-order-solution} and
\S\ref{sec:Perturbed-Einstein-Equations}. 
The gauge invariant variables ${\cal H}_{ab}$ and ${\cal L}_{ab}$ have
six independent components, while the original metric perturbations
$h_{ab}$ and $l_{ab}$ have ten independent components.

%*********************************************\\

Next, we consider the $O(\epsilon\lambda)$ metric perturbation
$k_{ab}$, which transformed as 
\begin{eqnarray}
  {}^{{\cal Y}}k_{ab} - {}^{{\cal X}}k_{ab}
  &=& 
  {\pounds}_{{}^{(\epsilon)}\xi} {}^{{\cal X}}l_{ab}
  + {\pounds}_{{}^{(\lambda)}\xi} {}^{{\cal X}}h_{ab}
  \nonumber\\
  && \quad\quad
  + \left\{ {\pounds}_{{}^{(\epsilon\lambda)}\xi}
    + \frac{1}{2}
    {\pounds}_{{}^{(\epsilon)}\xi}{\pounds}_{{}^{(\lambda)}\xi} 
    + \frac{1}{2}
    {\pounds}_{{}^{(\lambda)}\xi}{\pounds}_{{}^{(\epsilon)}\xi} 
  \right\} \eta_{ab}
\end{eqnarray}
under the gauge transformation $\Phi={\cal X}^{-1}\circ{\cal Y}$.
Using the gauge variant parts ${}^{(\epsilon)}X_{b}$ and
${}^{(\lambda)}X_{b}$ of the $O(\epsilon)$ and $O(\lambda)$ 
metric perturbations, we first define the variable $\widehat{\cal K}_{ab}$ by 
\begin{eqnarray}
  \label{eq:calKhat-def}
  \widehat{\cal K}_{ab} 
  &:=& 
  k_{ab} 
  - {\pounds}_{{}^{(\epsilon)}X}l_{ab} 
  - {\pounds}_{{}^{(\lambda)}X}h_{ab}
  \nonumber\\
  && \quad
  + \frac{1}{2}\left\{
    {\pounds}_{{}^{(\epsilon)}X} {\pounds}_{{}^{(\lambda)}X} 
    + {\pounds}_{{}^{(\epsilon)}X} {\pounds}_{{}^{(\lambda)}X}
  \right\} \eta_{ab}.
\end{eqnarray}
We note that the tensor $\widehat{\cal K}_{ab}$ has ten independent
components, like the usual metric.
Further, under the gauge transformation 
$\Phi={\cal X}^{-1}\circ{\cal Y}$, $\widehat{\cal K}_{ab}$ is
transformed as 
\begin{eqnarray}
  \label{eq:widehat-calK-gauge-trans}
  {}^{{\cal Y}}\widehat{\cal K}_{ab} 
  - {}^{{\cal X}}\widehat{\cal K}_{ab} &=& {\pounds}_{\sigma}\eta_{ab},
\end{eqnarray}
where 
\begin{equation}
  \sigma^{a} := {}^{(\epsilon\lambda)}\xi^{a} 
  + \frac{1}{2}[{}^{(\lambda)}\xi,{}^{(\epsilon)}X]^{a} 
  + \frac{1}{2}[{}^{(\epsilon)}\xi,{}^{(\lambda)}X]^{a}.
\end{equation}
The gauge transformation (\ref{eq:widehat-calK-gauge-trans}) has
exactly the same form as the linear-order metric perturbations.
Then, using a procedure similar to that used to obtain the
decompositions (\ref{eq:epsilon-metric-decomp}) and
(\ref{eq:lambda-metric-decomp}), we can decompose 
$\widehat{\cal K}_{ab}$ as 
\begin{equation}
  \widehat{\cal K}_{ab} 
  =:
  {\cal K}_{ab} 
  + 2 \nabla_{(a} {}^{(\epsilon\lambda)}X_{b)},
  \label{eq:epsilonlambda-metric-decomp}
\end{equation}
where the variables ${\cal K}_{ab}$ and ${}^{(\epsilon\lambda)}X_{a}$ 
are transformed as 
\begin{eqnarray}
  {}^{{\cal Y}}{\cal K}_{ab} - {}^{{\cal X}}{\cal K}_{ab} = 0,
  \quad
  {}^{{\cal Y}(\epsilon\lambda)}X_{a}-{}^{{\cal X}(\epsilon\lambda)}X_{a} =
  \sigma_{a}
\end{eqnarray}
under the gauge transformation $\Phi={\cal X}^{-1}\circ{\cal Y}$.
Thus, we can extract the gauge invariant part ${\cal K}_{ab}$ from the
$O(\epsilon\lambda)$ metric perturbation $k_{ab}$.

%*********************************************\\

Finally, we define the gauge invariant variable for the displacement
perturbation $\xi_{1}^{\alpha}$ and the energy density perturbation
$\delta\rho$.
The procedure to construct the corresponding gauge invariant variables
for any perturbation is shown in the forthcoming
paper.\cite{kouchan-two-parameters}
The displacement perturbation $\xi_{1}^{\alpha}$ is the $O(\epsilon)$
perturbation of a scalar function $\xi^{\alpha}$ and is transformed
as 
\begin{equation}
  {}^{{\cal Y}}\xi_{1}^{\alpha} - {}^{{\cal X}}\xi_{1}^{\alpha}
  = {\pounds}_{{}^{(\epsilon)}\xi}\xi^{\alpha} =
  (\zeta^{\alpha})_{a}{}^{(\epsilon)}\xi^{a}
  \label{eq:gauge-trans-xi-1-alpha}
\end{equation}
under the gauge transformation $\Phi={\cal X}^{-1}\circ{\cal Y}$.
On the other hand, the energy density perturbation $\delta\rho$ is the
$O(\epsilon\lambda)$ perturbation of the $O(\lambda)$ perturbation
$\rho$. 
We note that the corresponding background value and the $O(\epsilon)$
perturbation are trivial. 
Then, under the gauge transformation $\Phi={\cal X}^{-1}\circ{\cal Y}$,
$\delta\rho$ is transformed as 
\begin{equation}
  {}^{{\cal Y}}\delta\rho - {}^{{\cal X}}\delta\rho
  = {\pounds}_{{}^{(\epsilon)}\xi} {}^{{\cal X}}\rho
  = {}^{(\epsilon)}\xi^{a}\nabla_{a} {}^{{\cal X}}\rho.
  \label{eq:gauge-trans-delta-rho}
\end{equation}
Inspecting the gauge transformation rules
(\ref{eq:gauge-trans-xi-1-alpha}) and
(\ref{eq:gauge-trans-delta-rho}), we define the variables
\begin{eqnarray}
  \widehat{\Sigma} := \delta \rho 
  - {}^{(\epsilon)}X^{a} \nabla_{a} \rho,
  \quad\quad
  \widehat{V}_{a} := \gamma_{\alpha\beta}(\zeta^{\alpha})_{a} \xi^{\beta}_{1}
  - \gamma_{ab} {}^{(\epsilon)}X^{b},
  \label{eq:Sigmahat-Vhat-def}
\end{eqnarray}
where $\widehat{\Sigma}$ and $\widehat{V}_{a}$ are the gauge
invariant variables corresponding to the energy density
perturbation and the perturbative displacement of the string,
respectively. 
Using these gauge invariant variables, we derive the perturbative
Einstein equations of orders $\epsilon$, $\lambda$ and
$\epsilon\lambda$.

%*********************************************\\

%%%%%%%%%%%%%%%%%%%%%%%%%%%%%%%%%%%%%%%%%%%%%%%%%%%%%%%%%%%%%%%%%%%%%%
\section{$O(\epsilon)$ and $O(\lambda)$ solutions}
\label{sec:epsilon-order-lambda-order-solution}
%%%%%%%%%%%%%%%%%%%%%%%%%%%%%%%%%%%%%%%%%%%%%%%%%%%%%%%%%%%%%%%%%%%%%%

%*********************************************\\

Here, we derive the $O(\epsilon)$ and $O(\lambda)$ solutions to the
$O(\epsilon)$ and $O(\lambda)$ Einstein equations, respectively. 
To derive the perturbed Einstein equation for each order, the formulae
for the curvature expansion given in Appendix
\ref{sec:Curvature-Appendix} are useful. 
These can be derived by straightforward calculations.

%*********************************************\\

%%%%%%%%%%%%%%%%%%%%%%%%%%%%%%%%%%%%%%%%%%%%%%%%%%%%%%%%%%%%%%%%%%%%%%
\subsection{$O(\epsilon)$ solutions}
\label{sec:epsilon-order-solution}

%*********************************************\\

First, we consider the $O(\epsilon)$ solutions.
Since there is no $O(\epsilon)$ term in the energy-momentum tensor
(\ref{eq:energy-momentum-perturbation}), the $O(\epsilon)$ Einstein
equations are the linearized vacuum equations.
Explicitly, the $O(\epsilon)$ Einstein equations
$\left. \frac{\partial}{\partial\epsilon}G_{a}^{\;\;b}
\right|_{\lambda=\epsilon=0}=0$ are given by 
\begin{eqnarray}
  \label{eq:epsilon-Einstein-eq}
  && \nabla_{[a}^{} {\cal H}_{c]b}^{\;\;\;\;\;c}  
  - \frac{1}{2}\eta_{ab}
  \nabla_{[c}^{} {\cal H}_{d]}^{\;\;\;cd}  
  = 0,
%  \quad
  \\
  && {\cal H}_{ab}^{\;\;\;\;c} :=
  \nabla_{(a}^{} {\cal H}_{b)}^{\;\;c} - \frac{1}{2}
  \nabla^{c} {\cal H}_{ab}.
  \label{eq:calHabc-def}
\end{eqnarray}
At this order, the infinite string may oscillate but, it does not
produce a gravitational field. 
The analysis for the $O(\epsilon)$ perturbations are completely
parallel to the treatment of cylindrical and stationary perturbations
at $O(\lambda)$ and that of the dynamical perturbations at
$O(\epsilon\lambda)$.
Through this analysis, we easily find that the gauge invariant metric
perturbation ${\cal H}_{ab}$ describes only the free propagation of
gravitational waves, which has nothing to do with the string
oscillations.
We stipulate that there is no such gravitational wave.
Hence, as the $O(\epsilon)$ solution, we obtain 
${\cal H}_{ab}=0$, and $h_{ab}$ describes the pure gauge solution:
\begin{eqnarray}
  \label{eq:epsilon-metric-solution}
  h_{ab} = 2 \nabla_{(a}{}^{(\epsilon)}X_{b)}.
\end{eqnarray}

%*********************************************\\

%%%%%%%%%%%%%%%%%%%%%%%%%%%%%%%%%%%%%%%%%%%%%%%%%%%%%%%%%%%%%%%%%%%%%%
\subsection{$O(\lambda)$ solutions}
\label{sec:lambda-order-solution}

%*********************************************\\

Next, we consider the $O(\lambda)$ solutions.
Explicitly, the $O(\lambda)$ Einstein equations
$\left.\frac{\partial}{\partial\lambda} 
  \left( G_{a}^{\;\;b} - 8\pi G T_{a}^{\;\;b} \right)
\right|_{\lambda=\epsilon=0} = 0$ are given by 
\begin{eqnarray}
  \label{eq:lambda-Einstein-eq}
  && - 2 \left(
    \nabla_{[a}^{} {\cal L}_{c]b}^{\;\;\;\;\;c}  
    - \frac{1}{2}\eta_{ab}
    \nabla_{[c}^{} {\cal L}_{d]}^{\;\;\;cd}  
  \right)
  = - 8\pi G \rho q_{ab},
  \\
%  \quad
  \label{eq:calLabc-def}
  && {\cal L}_{ab}^{\;\;\;\;c} :=
  \nabla_{(a} {\cal L}_{b)}^{\;\;c} - \frac{1}{2}
  \nabla^{c} {\cal L}_{ab}.
\end{eqnarray}
At this order, an infinite string is static, without oscillations,
and it produces a static gravitational potential.  
In addition to this static potential, the linearized Einstein
equations of $O(\lambda)$ also describe the free propagation of
gravitational waves,  as at $O(\epsilon)$.
The derivation of this free propagation of gravitational waves
is completely parallel to the analysis for the dynamical perturbations
at $O(\epsilon\lambda)$.
Here, we present only the treatment of cylindrical and stationary
perturbations and the derivation of the static gravitational potential
produced by the infinite string.
The treatment given here also demonstrates the procedure to
obtain the decomposition (\ref{eq:epsilon-metric-decomp}), 
(\ref{eq:lambda-metric-decomp}), and
(\ref{eq:epsilonlambda-metric-decomp}) and thereby extract the gauge
invariant part from the metric perturbations $h_{ab}$, $l_{ab}$ and 
$\widehat{\cal K}_{ab}$ in the case of cylindrical and stationary
perturbations.

%*********************************************\\

The cylindrical and stationary perturbations are characterized by 
\begin{equation}
  {\cal D}_{a}Q = 0,
\end{equation}
where $Q$ formally represents all perturbative variables of
$O(\lambda)$.
Though Eq.~(\ref{eq:lambda-Einstein-eq}) is given in terms of
gauge invariant variables, it is convenient to start from the
$O(\lambda)$ metric perturbation $l_{ab}$ itself.
Since $\nabla_{a}$ is the covariant derivative associated with the
flat metric, we easily check that
\begin{equation}
  L_{ab}^{\;\;\;\;c} := \nabla_{(a} l_{b)}^{\;\;c} - \frac{1}{2}
  \nabla^{c} l_{ab} = {\cal L}_{ab}^{\;\;\;\;c} +
  \nabla_{a}\nabla_{b}{}^{(\lambda)}X^{c}. 
\end{equation}
Using the tensor $L_{ab}^{\;\;\;\;c}$, the right-hand side of
Eq.~(\ref{eq:lambda-Einstein-eq}), which is just 
$\left.\frac{\partial}{\partial\lambda}G_{a}^{\;\;b}\right|_{\lambda=\epsilon=0}$,
is given by 
\begin{eqnarray}
  \label{eq:lambda-Einstein-eq-2}
  && - 2 \left(
    \nabla_{[a}^{} {\cal L}_{c]b}^{\;\;\;\;\;c}  
    - \frac{1}{2}\eta_{ab}
    \nabla_{[c}^{} {\cal L}_{d]}^{\;\;\;cd}  
  \right)
  =
  - 2 \left(
    \nabla_{[a}^{} L_{c]b}^{\;\;\;\;\;c}  
    - \frac{1}{2}\eta_{ab}
    \nabla_{[c}^{} L_{d]}^{\;\;\;cd}  
  \right).
\end{eqnarray}
Under the gauge transformation $\Phi={\cal X}^{-1}\circ{\cal Y}$, 
the $O(\lambda)$ metric perturbation $l_{ab}$ is transformed as in 
Eq.~(\ref{eq:lambda-metric-gauge-trans}).
For cylindrical and stationary perturbations, the
transformation rule (\ref{eq:lambda-metric-gauge-trans}) is given by 
\begin{eqnarray}
  \label{eq:14.70}
  {}^{{\cal Y}}l_{\alpha\beta} 
  - {}^{{\cal X}}l_{\alpha\beta} 
  &=& D_{\alpha}{}^{(\lambda)}\xi_{\beta} + D_{\beta}{}^{(\lambda)}\xi_{\alpha}, \\
  \label{eq:14.71}
  {}^{{\cal Y}}l_{i\alpha} 
  - {}^{{\cal X}}l_{i\alpha} 
  &=& D_{\alpha}{}^{(\lambda)}\xi_{i}, \\
  \label{eq:14.72}
  {}^{{\cal Y}}l_{ij} - {}^{{\cal X}}l_{ij} &=& 0.
\end{eqnarray}

%*********************************************\\

For cylindrical and stationary perturbations, the components of
the $O(\lambda)$ Einstein equations (\ref{eq:lambda-Einstein-eq-2})
are given by 
\begin{eqnarray}
  &&
  - D_{\alpha}D^{\beta}\left( l_{i}^{\;\;i} 
    + l_{\gamma}^{\;\;\gamma}\right) 
  + D^{\gamma} D_{\alpha} l_{\gamma}^{\;\;\beta} 
  + D^{\gamma} D^{\beta} l_{\alpha\gamma} 
  - \Delta l_{\alpha}^{\;\;\beta} 
  \nonumber\\
  && \quad\quad
  + \delta_{\alpha}^{\;\;\beta} \left( 
    \Delta
    (l_{\delta}^{\;\;\delta} + l_{i}^{\;\;i}) 
    - D^{\gamma} D^{\delta} l_{\delta\gamma} 
  \right) = 0,
  \label{eq:14.66}
  \\
  && D^{\gamma} D_{\alpha} l_{\gamma}^{\;\;i} 
  - \Delta l_{\alpha}^{\;\;i}
  = 0,
  \label{eq:14.67}
  \\
  && 
  - \Delta l_{i}^{\;\;j} 
  + \delta_{i}^{\;\;j} \left( 
    \Delta (l_{\delta}^{\;\;\delta} + l_{l}^{\;\;l}) 
    - D^{\gamma} D^{\delta}l_{\gamma\delta} 
  \right) = - 16\pi G \delta_{i}^{\;\;j} \rho,
  \label{eq:14.69}
\end{eqnarray}
where $\delta_{\alpha}^{\;\;\beta}$ and $\delta_{i}^{\;\;j}$ are
the two-dimensional Kronecker delta, and 
$\Delta := D^{c}D_{c} = D^{\gamma}D_{\gamma}$ is the
Laplacian associated with the metric $\gamma_{ab}$.

%*********************************************\\

From the trace of (\ref{eq:14.66}) and  (\ref{eq:14.69}), we obtain
\begin{eqnarray}
  \label{eq:trace-of-D.18}
  && \Delta l_{i}^{\;\;i} = 0, \\
  \label{eq:trace-of-D.21}
  && \Delta l_{\delta}^{\;\;\delta}
  - D^{\gamma} D^{\delta} l_{\gamma\delta} = - 16\pi G\rho.
\end{eqnarray}
Further, substituting (\ref{eq:trace-of-D.18}) and
(\ref{eq:trace-of-D.21}) into (\ref{eq:14.69}), we obtain 
\begin{equation}
  \label{eq:nontrace-of-D.21}
  \Delta l_{i}^{\;\;j} = 0.
\end{equation}
Here, we impose the condition that there is no singular behavior of
$l_{i}^{\;\;j}$ on ${\cal M}_{2}$. 
Then (\ref{eq:nontrace-of-D.21}) implies that $l_{i}^{\;\;j}$ should
be constant on ${\cal M}_{2}$. 
This constant can be removed through a scale transformation of the
coordinates $\sigma^{i}$. 
Then we may choose $l_{i}^{\;\;j} = 0$ without loss of generality.

%*********************************************\\

Next, we consider Eq.~(\ref{eq:14.67}).
Here, we decompose $l_{\alpha i}$ as
\begin{equation}
  \label{eq:decomposition-of-lalphai}
  l_{\alpha i} = D_{\alpha} f_{i} 
  + {\cal E}_{\alpha\beta} D^{\beta} g_{i},
\end{equation}
where ${\cal E}_{\alpha\beta}$ is the component of the two-dimensional
totally antisymmetric tensor defined by
\begin{equation}
  \label{eq:cal-Epsilon-ab-def}
  {\cal E}_{ab} := 
  {\cal E}_{\alpha\beta}(d\xi^{\alpha})_{a}(d\xi^{\alpha})_{b} 
  := (dx)_{a}(dy)_{b} - (dy)_{a}(dx)_{b}.
\end{equation}
From (\ref{eq:14.71}), it is clear that $f_{i}$ represents the gauge freedom.
Substituting the decomposition (\ref{eq:decomposition-of-lalphai})
into Eq.~(\ref{eq:14.67}), we obtain 
\begin{equation}
  {\cal E}_{\alpha\beta} D^{\beta} \Delta g_{i} = 0.
\end{equation}
This shows that $\Delta g_{i}$ is a constant on ${\cal M}_{2}$.
If this constant is non-vanishing, then $l_{\alpha i}$
proportional to the distance from the axis of the cylindrical
symmetry and diverges at infinity (i.e. in the region infinitely far
from the string).  
Here, we impose the condition that there is no divergence of the
perturbative metric. In this case, $g_{i}$ is a solution to the
two-dimensional Laplace equation on the flat space.
We can easily check that if $g_{i}$ is a solution to the
two-dimensional Laplace equation, there is a function $\tilde{f}_{i}$
such that
\begin{equation}
  {\cal E}_{\alpha\beta} D^{\beta} g_{i} = D_{\alpha} \tilde{f}_{i}.
\end{equation}
Therefore, $g_{i}$ is also regarded as representing the gauge
freedom. 
Thus, we obtain the solution to the equation (\ref{eq:14.67})
\begin{equation}
  l_{\alpha i} = D_{\alpha} f_{i},
\end{equation}
which represents only the gauge freedom of perturbations.

%*********************************************\\

Finally, we consider the components $l_{\alpha\beta}$. 
We decompose $l_{\alpha\beta}$ as 
\begin{equation}
  \label{eq:calKalphabeta-decomp}
  l_{\alpha\beta} = \frac{1}{2}\gamma_{\alpha\beta}\Xi
  + D_{\alpha}D_{\beta}F + {\cal A}_{\alpha\beta}G,
\end{equation}
where the derivative operator ${\cal A}_{\alpha\beta}$ is defined by 
\begin{equation}
  \label{eq:cal-A-ab-def}
  {\cal A}_{\alpha\beta}:={\cal E}_{\gamma(\alpha}D_{\beta)}D^{\gamma}.
\end{equation}
We also decompose the generator ${}^{(\lambda)}\xi^{a}$ of the gauge
transformation as
\begin{equation}
  {}^{(\lambda)}\xi_{\alpha} = D_{\alpha} f + {\cal E}_{\alpha\beta}D^{\beta}g.
\end{equation}
Using this decomposition, the gauge transformation
(\ref{eq:14.70}) is given by  
\begin{equation}
  {}^{{\cal Y}}l_{\alpha\beta} - {}^{{\cal X}}l_{\alpha\beta} 
  = D_{\alpha}D_{\beta} f + 2 {\cal A}_{\alpha\beta}g.
\end{equation}
This shows that the functions $F$ and $G$ in the decomposition
(\ref{eq:calKalphabeta-decomp}) represent the gauge freedom, and
only the function $\Xi$ is gauge invariant.
By substitution of the decomposition
(\ref{eq:calKalphabeta-decomp}) and $l_{i}^{\;\;j}=0$
into (\ref{eq:14.66}), we can easily check that (\ref{eq:14.66})
is trivial. 
Further, Eq.~(\ref{eq:trace-of-D.21}) with the decomposition 
(\ref{eq:calKalphabeta-decomp}) gives
\begin{equation}
  \label{eq:lambda-order-Einstein-eq}
  \Delta \Xi = - 32 \pi G \rho.
\end{equation}

%*********************************************\\

Since $\Delta$ is the two-dimensional Laplacian, for a given 
function $\rho$, the solution $\Xi$ to
Eq.~(\ref{eq:lambda-order-Einstein-eq}) can be obtained by using usual  
Green function of the differential operator $\Delta$ on 
${\cal M}_{2}$. 
Further, we can easily see that the solution to
Eq.~(\ref{eq:lambda-order-Einstein-eq}) has a logarithmic divergence
at infinity.  
However, this logarithmic divergence can be interpreted as the
deficit angle produced by the string\cite{Vilenkin_Shellard}.
For this reason, we do not treat this divergence, seriously, and here we
regard this logarithmic divergence as representing the asymptotically
conical structure of the full spacetime $({\cal M},g_{ab})$.
Thus, using the solution to Eq.~(\ref{eq:lambda-order-Einstein-eq}),
the gauge invariant metric perturbation ${\cal L}_{ab}$ is given by  
\begin{eqnarray}
  {\cal L}_{ab} = \frac{1}{2} \gamma_{ab} \Xi.
\end{eqnarray}
Including the gauge freedom, we obtain the $O(\lambda)$ 
metric $l_{ab}$ as the solution to the Einstein equations
of $O(\lambda)$
\begin{eqnarray}
  \label{eq:lambda-metric-solution}
  l_{ab} = \frac{1}{2} \gamma_{ab} \Xi + 2 \nabla_{(a}{}^{(\lambda)}X_{b)},
\end{eqnarray}
at least for cylindrical and stationary perturbations.
With analysis completely parallel to that used for the dynamical
perturbations of $O(\epsilon\lambda)$, we can see that
(\ref{eq:lambda-metric-solution}) is the general solution to the
$O(\lambda)$ Einstein equations (\ref{eq:lambda-Einstein-eq-2}) if we 
stipulate that there is no free propagation of gravitational waves,
which have nothing to do with the string oscillations.

%*********************************************\\

When $\rho=0$, we note that $\Xi$ must be constant on ${\cal M}_{2}$
if we impose the metric perturbation $l_{ab}$ has no divergence in
${\cal M}_{2}$, except for the logarithmic divergence at infinity. 
This constant can be eliminated through a the conformal scale
transformation of the coordinates $\xi^{\alpha}$, and we may choose
$\Xi=0$ without loss of generality.
This implies that $O(\epsilon)$ perturbations do not include
cylindrical and stationary solutions, except for that representing the
gauge freedom.
Further, this also holds in $O(\epsilon\lambda)$ if $\delta\rho$
in (\ref{eq:energy-momentum-perturbation}) does not include any
stationary components.

%*********************************************\\

%%%%%%%%%%%%%%%%%%%%%%%%%%%%%%%%%%%%%%%%%%%%%%%%%%%%%%%%%%%%%%%%%%%%%%
\section{$O(\epsilon\lambda)$ Einstein equations}
\label{sec:Perturbed-Einstein-Equations}
%%%%%%%%%%%%%%%%%%%%%%%%%%%%%%%%%%%%%%%%%%%%%%%%%%%%%%%%%%%%%%%%%%%%%%

%*********************************************\\

In this section, we consider dynamical perturbations of
$O(\epsilon\lambda)$.
We derive the $O(\epsilon\lambda)$ Einstein equations using the mode
expansions in the harmonics defined in Appendix
\ref{sec:harmonics-Appendix}.  
The treatment of the dynamical perturbation in this section also shows
that the procedure for the decomposition (\ref{eq:epsilon-metric-decomp}), 
(\ref{eq:lambda-metric-decomp}) and
(\ref{eq:epsilonlambda-metric-decomp}) and the resulting extraction of
the gauge invariant part from the metric perturbations $h_{ab}$,
$l_{ab}$ and $\widehat{\cal K}_{ab}$ in the case of dynamical
perturbations.

%*********************************************\\

As mentioned in \S\ref{sec:epsilon-order-solution}, we have imposed
the condition that the free propagation of gravitational waves that
have nothing to do with the string oscillations ; i.e., 
${\cal H}_{ab}=0$. 
Then, using the formula
(\ref{eq:ep-lam-order-Ricci-curvature-without-gauge-fix}), we can
derive the $O(\epsilon\lambda)$ Einstein equation 
$\left.\frac{\partial^{2}}{\partial\lambda\partial\epsilon} 
\left( G_{a}^{\;\;b} - 8\pi G T_{a}^{\;\;b} \right)
\right|_{\lambda=\epsilon=0} = 0$ 
in terms of the gauge invariant variables defined by
Eqs.~(\ref{eq:epsilonlambda-metric-decomp}) and (\ref{eq:Sigmahat-Vhat-def}): 
\begin{eqnarray}
  \label{eq:calKabc-def}
  && {\cal K}_{ab}^{\;\;\;\;c} :=
  \nabla_{(a} {\cal K}_{b)}^{\;\;c} - \frac{1}{2}
  \nabla^{c} {\cal K}_{ab}, \\
  \label{eq:epsilon-lambda-Einstein-eq}
  && - 2 \left(
    \nabla_{[a} {\cal K}_{c]b}^{\;\;\;\;\;c}  
    - \frac{1}{2}\eta_{ab}
    \nabla_{[c} {\cal K}_{d]}^{\;\;\;cd}  
  \right)
  = 8\pi G \left(
    - q_{ab} \widehat{\Sigma}
    + 2 \rho q_{c(a} \nabla^{c} \widehat{V}_{b)}
  \right).
\end{eqnarray}

%*********************************************\\

To evaluate the Einstein equations
(\ref{eq:epsilon-lambda-Einstein-eq}), we must carry out the
decomposition (\ref{eq:epsilonlambda-metric-decomp}) by inspecting the  
gauge transformation rules (\ref{eq:widehat-calK-gauge-trans}).
This is given by 
\begin{eqnarray}
  \label{eq:widehat-calKalphabeta-gauge-trans}
  {}^{{\cal Y}}\widehat{\cal K}_{\alpha\beta} 
  - {}^{{\cal X}}\widehat{\cal K}_{\alpha\beta} 
  &=& D_{\alpha} \zeta_{\beta} + D_{\beta} \zeta_{\alpha}, \\
  \label{eq:widehat-calKalphai-gauge-trans}
  {}^{{\cal Y}}\widehat{\cal K}_{\alpha i} 
  - {}^{{\cal X}}\widehat{\cal K}_{\alpha i} 
  &=& D_{\alpha} \zeta_{i} + {\cal D}_{i} \zeta_{\alpha}, \\
  \label{eq:widehat-calKij-gauge-trans}
  {}^{{\cal Y}}\widehat{\cal K}_{ij} 
  - {}^{{\cal X}}\widehat{\cal K}_{ij} 
  &=& {\cal D}_{i} \zeta_{j} + {\cal D}_{j} \zeta_{i}.
\end{eqnarray}
To accomplish the decomposition (\ref{eq:epsilonlambda-metric-decomp})
for dynamical perturbations, it is convenient to use the expansion in
harmonic functions summarized in Appendix \ref{sec:harmonics-Appendix}.
The procedure to carry out the decompositions
(\ref{eq:epsilon-metric-decomp}) and (\ref{eq:lambda-metric-decomp})
is completely the same.
As mentioned in Appendix \ref{sec:harmonics-Appendix}, we should
distinguish perturbative modes for which $\kappa$ defined by
(\ref{eq:kappa-def}) vanishes from those for which it does not. 
The $\kappa=0$ modes correspond to the perturbative modes that 
propagate along the string, and the $\kappa\neq 0$ modes are the
other dynamical modes.
Though the cylindrical and stationary perturbative modes also
exist in the present case, as in $O(\lambda)$ perturbations, we do not
treat these modes here. 
Clearly, these modes have nothing to do with the dynamics of the
string.
This can be easily seen from treatments that are completely the same
as that in \S\ref{sec:lambda-order-solution}.

%*********************************************\\

%%%%%%%%%%%%%%%%%%%%%%%%%%%%%%%%%%%%%%%%%%%%%%%%%%%%%%%%%%%%%%%%%%%%%
\subsection{$\kappa\neq 0$ mode perturbation}
\label{sec:kappaneq0-mode-perturbation-variables}

%*********************************************\\

First, we consider the case in which the all perturbative variables
satisfy the conditions 
\begin{equation}
  {\cal D}_{a}Q \neq 0, \quad  {\cal D}^{a}{\cal D}_{a}Q \neq 0,
\end{equation}
where $Q$ formally represents all the perturbative variables,
$\widehat{\cal K}_{ab}$, $\widehat{\Sigma}$, $\widehat{V}^{a}$ and
$\zeta^{a}$.
Using the $\kappa\neq0$ mode harmonics introduced in
Appendix \ref{sec:kappaneq0-mode-harmonics}, these perturbative
variables are expanded as follows:
\begin{eqnarray}
  \label{eq:calKalphabeta_decomp_def_kneq0}
  \widehat{\cal K}_{\alpha\beta} &=:& \int f_{\alpha\beta} S, \quad\quad
  \widehat{\cal K}_{\alpha i} =: \int \left[f_{\alpha(o1)} V_{(o1)i} 
    + f_{\alpha(e1)} V_{(e1)i}\right],\\
  \label{eq:calKij_decomp_def_kneq0}
  \widehat{\cal K}_{ij} &=:& \int \left[f_{(o2)} T_{(o2)ij} 
    + f_{(e0)} T_{(e0)ij} + f_{(e2)} T_{(e2)ij}\right], \\
  \label{eq:zetaalpha_decomp_def_kneq0}
  \zeta_{\alpha} &=:& \int \iota_{\alpha} S, \quad\quad
  \zeta_{i} =: \int \left[\iota_{(o)} V_{(o)i} 
    + \iota_{(e)} V_{(e)i}\right], \\
  \label{eq:sigma_decomp_def_kneq0}
  \widehat{\Sigma} &=:& - \frac{1}{32\pi G} \int \Sigma S, \quad\quad
  \rho \widehat{V}_{\alpha} =: \frac{1}{16\pi G} \int V_{\alpha} S,
  \quad\quad \rho \widehat{V}_{i} = 0,
\end{eqnarray}
where the measure $\int$ of the harmonic analyses is given by 
$\int = \int d^{2}k$.
Throughout this subsection, we use the measure $\int$ with this
meaning. 
All expansion coefficients defined by
(\ref{eq:calKalphabeta_decomp_def_kneq0})--(\ref{eq:sigma_decomp_def_kneq0})
are tensors on $({\cal M}_{2},\gamma_{ab})$.

%*********************************************\\

By inspecting the gauge transformation rules of the coefficients,
we easily find gauge invariant combinations. 
In terms of the mode coefficients defined by 
(\ref{eq:calKalphabeta_decomp_def_kneq0})--(\ref{eq:zetaalpha_decomp_def_kneq0})
the gauge transformation rules
(\ref{eq:widehat-calKalphabeta-gauge-trans})--(\ref{eq:widehat-calKij-gauge-trans})
are given by 
\begin{eqnarray}
  \label{eq:delta_G_f_ab-kneq0}
  && {}^{{\cal Y}}f_{\alpha\beta} - {}^{{\cal X}}f_{\alpha\beta} 
  = D_{\alpha}\iota_{\beta} + D_{\beta}\iota_{\alpha}, \\
  \label{eq:delta_G_f_ae1-kneq0}
  && {}^{{\cal Y}}f_{\alpha(e1)} - {}^{{\cal X}}f_{\alpha(e1)} 
  = D_{\alpha}\iota_{(e)} + \iota_{\alpha}, \\
  \label{eq:delta_G_f_ae0-kneq0}
  && {}^{{\cal Y}}f_{(e0)} - {}^{{\cal X}}f_{(e0)} 
  = 2 \kappa^{2} \iota_{(e)}, \\
  \label{eq:delta_G_f_ae2-kneq0}
  && {}^{{\cal Y}}f_{(e2)} - {}^{{\cal X}}f_{(e2)} 
  = 2 \iota_{(e)}, \\
  \label{eq:delta_G_f_ao1-kneq0}
  && {}^{{\cal Y}}f^{\alpha}_{(o1)} - {}^{{\cal X}}f^{\alpha}_{(o1)} 
  = D^{\alpha}\iota_{(o)}, \\
  \label{eq:delta_G_f_o2-kneq0}
  && {}^{{\cal Y}}f_{(o2)} - {}^{{\cal X}}f_{(o2)} 
  = 2 \iota_{(o)}.
\end{eqnarray}
From these transformation rules, we see that there exist six
independent gauge-invariant quantities. 
First, from Eqs.~(\ref{eq:delta_G_f_ao1-kneq0}),
(\ref{eq:delta_G_f_o2-kneq0}) and (\ref{eq:delta_G_f_ae2-kneq0}), we
see that there exist two independent gauge invariant quantities
defined by 
\begin{equation}
  \label{eq:F_a_def-kneq0}
  F_{\alpha} := f_{\alpha(o1)} - \frac{1}{2}D_{\alpha}f_{(o2)}.
\end{equation}
Second, we define 
\begin{equation}
  \label{eq:F_def-kneq0}
  F := f_{(e0)} - \kappa^2 f_{(e2)}.
\end{equation}
The gauge transformations (\ref{eq:delta_G_f_ae0-kneq0}) reveal that
$F$ is a gauge invariant quantity. 
To find the remaining gauge invariants, we define the variable 
\begin{equation}
  \label{eq:Z_def-kneq0}
  Z^{\alpha} := f^{\alpha}_{(e1)} - \frac{1}{2} D^{\alpha}f_{(e2)}.
\end{equation}
From Eqs.~(\ref{eq:delta_G_f_ae1-kneq0}) and
(\ref{eq:delta_G_f_ae2-kneq0}), $Z^{\alpha}$ is transformed as 
\begin{equation}
  \label{eq:delta_G_Z_a-kneq0}
  {}^{{\cal Y}}Z^{\alpha} - {}^{{\cal X}}Z^{\alpha} = \iota^{\alpha}.
\end{equation}
Hence, we can easily obtain the gauge-invariant combinations
\begin{eqnarray}
  \label{eq:F_ab-kneq0}
  F_{\alpha\beta} := f_{\alpha\beta} - 2 D_{(\alpha}Z_{\beta)}.
\end{eqnarray}
From the gauge transformation rules
(\ref{eq:delta_G_f_ab-kneq0}) and (\ref{eq:delta_G_Z_a-kneq0})
and this definition, we easily check that $F_{\alpha\beta}$ is
gauge invariant.

%*********************************************\\

Thus, we obtain the six gauge invariant variables $F$, $F_{\alpha}$
and $F_{\alpha\beta}$. 
We note that the other four components are gauge variant. 
In terms of these variables, the metric perturbations 
$\widehat{\cal K}_{ab}$ are given by  
\begin{eqnarray}
  \label{hab_decomp_def_kneq0-with-GI}
  \widehat{\cal K}_{\alpha\beta} &=& \int \left(F_{\alpha\beta} 
    + 2 D_{(\alpha}Z_{\beta)}\right) S, \\ 
  \label{hap_decomp_def_kneq0-with-GI}
  \widehat{\cal K}_{\alpha i} &:=& \int \left[
    \left(  F_{\alpha} + \frac{1}{2}D_{\alpha}f_{(o2)}\right) V_{(o1)i} 
    + \left( Z_{\alpha} 
      + \frac{1}{2} D_{\alpha}f_{(e2)} \right) V_{(e1)i}\right],\\
  \label{hpq_decomp_def_kneq0-with-GI}
  \widehat{\cal K}_{ij} &:=& \int \left[f_{(o2)} T_{(o2)ij} 
    + (F + k^2 f_{(e2)}) T_{(e0)ij} + f_{(e2)} T_{(e2)ij}\right].
\end{eqnarray}
This implies that we have accomplished the decomposition of 
$\widehat{\cal K}_{ab}$ into ${\cal K}_{ab}$ and
$2\nabla_{(a}{}^{(\epsilon\lambda)}X_{b)}$ in the case of $\kappa\neq 0$
modes, obtaining
\begin{eqnarray}
  \label{eq:calKab_kneq0-GI}
  {\cal K}_{\alpha\beta} := \int F_{\alpha\beta} S, \quad 
  {\cal K}_{\alpha i} := \int F_{\alpha} V_{(o1)i}, \quad 
  {\cal K}_{ij} := \int F T_{(e0)ij}
\end{eqnarray}
and 
\begin{eqnarray}
  \label{eq:eplamX-gauge-variant-rela-kneq0}
  {}^{(\epsilon\lambda)}X_{\alpha} := \int Z_{\alpha} S, \quad
  {}^{(\epsilon\lambda)}X_{i} := \frac{1}{2} \int \left[
    f_{(o2)} V_{(o1)i} + f_{(e2)} V_{(e1)i}\right].
\end{eqnarray}

%*********************************************\\

Substituting Eqs.~(\ref{eq:sigma_decomp_def_kneq0}) and
(\ref{eq:calKab_kneq0-GI}) into (\ref{eq:epsilon-lambda-Einstein-eq}),
all components of the $O(\epsilon\lambda)$ Einstein equations for
$\kappa\neq 0$ modes are given by  
\begin{eqnarray}
  \label{eq:kneq-mode-equations-1}
  && (\Delta + \kappa^{2}) F = 0, \\
  \label{eq:kneq-mode-equations-2}
  && \left(\Delta + \kappa^{2}\right) F_{\alpha\beta}
  =  2 \left( D_{(\alpha}V_{\beta)} 
    - \frac{1}{2}\gamma_{\alpha\beta} D_{\gamma}V^{\gamma}\right), \\
  \label{eq:kneq-mode-equations-3}
  && D^{\gamma}F_{\alpha\gamma} - \frac{1}{2} D_{\alpha} F = V_{\alpha}, \\
  \label{eq:kneq-mode-equations-4}
  && F_{\gamma}^{\;\;\gamma} = 0, \\
  \label{eq:kneq-mode-equations-5}
  && D^{\gamma} V_{\gamma} + \frac{1}{2} \Sigma = 0 , \\
  \label{eq:kneq-mode-equations-6}
  && (\Delta + \kappa^{2}) F_{\alpha} = 0, \\
  \label{eq:kneq-mode-equations-7}
  && D_{\alpha}F^{\alpha} = 0. 
\end{eqnarray}
These equations are almost the same as the linearized Einstein equations
for $\kappa\neq 0$ modes obtained in our previous work\cite{kouchan-string-1}.
The differences are that there is no background curvature term
in Eq.~(\ref{eq:kneq-mode-equations-2}) and that the covariant derivative
$D_{a}$ is associated with the flat metric $\gamma_{ab}$, while
$D_{a}$ in our previous work is not.

%*********************************************\\

%%%%%%%%%%%%%%%%%%%%%%%%%%%%%%%%%%%%%%%%%%%%%%%%%%%%%%%%%%%%%%%%%%%%%
\subsection{$\kappa=0$ mode perturbation}
\label{sec:kappa=0-mode-perturbation-variables}

%*********************************************\\

Here, we consider the case in which the metric perturbation
$\widehat{\cal K}_{ab}$, the energy density perturbation
$\widehat{\Sigma}$, the displacement of the string world sheet
$\widehat{V}^{a}$, and the generator $\zeta^{a}$ of the gauge
transformations satisfy the equation
\begin{eqnarray}
  {\cal D}^{a}{\cal D}_{a} Q = 0, \quad
  {\cal D}_{a} Q \neq 0,
\end{eqnarray}
where $Q$ formally represents each of these perturbative variables.
In this case, using the $\kappa=0$ modes harmonics introduced in
Appendix \ref{sec:kappa=0-mode-harmonics}, these perturbative
variables are expanded as
\begin{eqnarray}
  \label{eq:calKalphabeta_decomp_def_k0}
  && \widehat{\cal K}_{\alpha\beta} =: \int f_{\alpha\beta}S, \quad\quad 
%  \label{eq:calKalphai_decomp_def_k0}
  \widehat{\cal K}_{\alpha i} =: \int \left[ f_{\alpha(l1)} V_{(l1)i} 
    + f_{\alpha(e1)} V_{(e1)i}\right],\\
  \label{eq:calKij_decomp_def_k0}
  && \widehat{\cal K}_{ij} =: \int \left[ f_{(l2)}
    T_{(l2)ij} + f_{(e0)} T_{(e0)ij} + f_{(e2)} T_{(e2)ij}\right], \\
  \label{eq:zetaalpha_decomp_def_k0}
  && \zeta_{\alpha} =: \int \iota_{\alpha}S, \quad\quad 
%  \label{eq:zetai_decomp_def_k0}
  \zeta_{i} =: \int 
  \left[ \iota_{(l)} V_{(l1)i} + \iota_{(e)} V_{(e1)i} \right], \\
  && \widehat{\Sigma} =: - \frac{1}{32\pi G} \int \Sigma S, \quad
  \rho \widehat{V}_{\alpha} =: \frac{1}{16\pi G} \int V_{\alpha} S, \quad
  \rho \widehat{V}_{i} = 0,
  \label{eq:sigma-Va-definition-k0}
\end{eqnarray}
where the measure $\int$ of the Fourier integration is given by
$\int = \sum_{\nu=\pm 1}\int d\omega$.
Throughout this subsection, we denote this measure by $\int$.
All expansion coefficients defined by
(\ref{eq:calKalphabeta_decomp_def_k0})--(\ref{eq:zetaalpha_decomp_def_k0})
are tensors on $({\cal M}_{2},\gamma_{ab})$.

%*********************************************\\

As in the case of $\kappa\neq 0$ mode perturbations, we can easily
find gauge invariant combinations for the metric perturbations by
inspecting the gauge transformation rules 
(\ref{eq:widehat-calKalphabeta-gauge-trans})--(\ref{eq:widehat-calKij-gauge-trans}) for the coefficients. 
In terms of the mode coefficients defined by 
(\ref{eq:calKalphabeta_decomp_def_k0})--(\ref{eq:zetaalpha_decomp_def_k0}),
the gauge transformation rules
(\ref{eq:widehat-calKalphabeta-gauge-trans})--(\ref{eq:widehat-calKij-gauge-trans})
are given by 
\begin{eqnarray}
  \label{eq:delta_G_f_ab}
  && {}^{{\cal Y}}f_{\alpha\beta} - {}^{{\cal X}}f_{\alpha\beta} 
  = D_{\alpha}\iota_{\beta} + D_{\beta}\iota_{\alpha}, \\
  \label{eq:delta_G_f_ae1}
  && {}^{{\cal Y}}f_{\alpha(e1)} - {}^{{\cal X}}f_{\alpha(e1)} 
  = D_{\alpha}\iota_{(e)} + \iota_{\alpha}, \\
  \label{eq:delta_G_f_al1}
  && {}^{{\cal Y}}f_{\alpha(l1)} - {}^{{\cal X}}f_{\alpha(l1)} 
  = D_{\alpha}\iota_{(l)}, \\
  \label{eq:delta_G_f_l2}
  && {}^{{\cal Y}}f_{(l2)} - {}^{{\cal X}}f_{(l2)} = 0, \\
  \label{eq:delta_G_f_ae0}
  && {}^{{\cal Y}}f_{(e0)} - {}^{{\cal X}}f_{(e0)} 
  = 4 \omega^{2} \iota_{(l)}, \\
  \label{eq:delta_G_f_ae2}
  && {}^{{\cal Y}}f_{(e2)} - {}^{{\cal X}}f_{(e2)} 
  = 2 \iota_{(e)}.
\end{eqnarray}

%*********************************************\\

From these transformation rules, we see that there exist six
independent gauge-invariant quantities. 
First, Eq.~(\ref{eq:delta_G_f_l2}) shows that $f_{(l2)}$ itself is a 
gauge invariant quantity:
\begin{equation}
  H := f_{(l2)}.
\end{equation}
Second, we define 
\begin{equation}
  H_{\alpha} := f_{\alpha(l1)} - \frac{1}{4\omega^{2}} D_{\alpha} f_{(e0)}. 
\end{equation}
The gauge transformation rules (\ref{eq:delta_G_f_al1}) and
(\ref{eq:delta_G_f_ae0}) imply that $H_{\alpha}$ is gauge invariant. 
To find the remaining gauge invariants, we define the variable  
\begin{equation}
  Z^{\alpha} := f^{\alpha}_{(e1)} - \frac{1}{2} D^{\alpha}f_{(e2)}.
\end{equation}
This definition, along with the gauge transformation rules
(\ref{eq:delta_G_f_ae1}) and (\ref{eq:delta_G_f_ae2}), shows that the
variable $Z^{\alpha}$ transforms as
\begin{equation}
  \label{eq:delta_G_Z_a}
  {}^{{\cal Y}}Z^{\alpha} - {}^{{\cal X}}Z^{\alpha} = \iota^{\alpha}.
\end{equation}
Using the variable $Z^{\alpha}$, we define
\begin{equation}
  H_{\alpha\beta} := f_{\alpha\beta} - 2 D_{(\alpha}Z_{\beta)}.
\end{equation}
From this definition and the gauge transformation rules
(\ref{eq:delta_G_f_ab}) and (\ref{eq:delta_G_Z_a}), we easily check
that $H_{\alpha\beta}$ is gauge invariant.

%*********************************************\\

From the above, we obtain the six gauge invariant variables $H$,
$H_{\alpha}$ and $H_{\alpha\beta}$. 
We note that the other four components are gauge variant. 
Using these variables, the metric perturbations 
$\widehat{\cal K}_{ab}$ are given by 
\begin{eqnarray}
  \label{eq:calKalphabeta_decomp_def_k0-with-GI}
  \widehat{\cal K}_{\alpha\beta} &=& \int
  \left(H_{\alpha\beta} + 2 D_{(\alpha}Z_{\beta)}\right) S, \\
  \label{eq:calKalphai_decomp_def_k0-with-GI}
  \widehat{\cal K}_{\alpha i} &=& \int \left[ 
    \left(H_{\alpha} 
      + \frac{1}{4\omega^{2}} D_{\alpha} f_{(e0)} \right) V_{(l1)i}  
    + \left(Z_{\alpha} 
      + \frac{1}{2} D_{\alpha}f_{(e2)}\right) V_{(e1)i} \right],\\
  \label{eq:calKij_decomp_def_k0-with-GI}
  \widehat{\cal K}_{ij} &=& \int \left[ H
    T_{(l2)ij} + f_{(e0)} T_{(e0)ij} + f_{(e2)} T_{(e2)ij}\right].
\end{eqnarray}
This implies that we have accomplished the decomposition of 
$\widehat{\cal K}_{ab}$ into ${\cal K}_{ab}$ and
$2\nabla_{(a}{}^{(\epsilon\lambda)}X_{b)}$ in the case of the
$\kappa=0$ modes, obtaining
\begin{eqnarray}
  \label{eq:calKab_k0-GI}
  {\cal K}_{\alpha\beta} := \int H_{\alpha\beta} S, \quad
  {\cal K}_{\alpha i} := \int H_{\alpha} V_{(l1)i}, \quad
  {\cal K}_{ij} := \int H T_{(l2)ij},
\end{eqnarray}
and 
\begin{eqnarray}
  \label{eq:eplamX-gauge-variant-rela-k0}
  {}^{(\epsilon\lambda)}X_{\alpha} := \int Z_{\alpha} S, \quad
  {}^{(\epsilon\lambda)}X_{i}
  := \frac{1}{2} \int \left[ 
    \frac{1}{2\omega^{2}} f_{(e0)} V_{(l1)i}  
      + f_{(e2)} V_{(e1)i} \right].
\end{eqnarray}

%*********************************************\\

Substituting Eqs.~(\ref{eq:sigma-Va-definition-k0}) and
(\ref{eq:calKab_k0-GI}) into (\ref{eq:epsilon-lambda-Einstein-eq}),
all components of the $O(\epsilon\lambda)$ Einstein equations
for the $\kappa=0$ modes are obtained as 
\begin{eqnarray}
  \label{eq:14.171}
  && \Delta H_{\alpha\beta} = 
  2 \left(D_{(\alpha}V_{\beta)} 
    - \frac{1}{2}\gamma_{\alpha\beta} D_{\gamma}V^{\gamma}\right), \\
  \label{eq:14.172}
  && D^{\gamma}H_{\gamma} + 2\omega^{2}H = 0, \\
  \label{eq:14.173}
  && D^{\gamma}H_{\gamma}^{\;\;\alpha} + 2 \omega^{2} H^{\alpha} 
  = V^{\alpha}, \\
  \label{eq:14.174}
  && \Delta H^{\alpha} = 0, \\
  \label{eq:14.175}
  && H_{\gamma}^{\;\;\gamma} = 0, \\
  \label{eq:14.176}
  && \Delta H = 0, \\ 
  \label{eq:14.177}
  && D_{\alpha}V^{\alpha} + \frac{1}{2}\Sigma = 0.
\end{eqnarray}
These equations are almost the same as the linearized Einstein
equations for the $\kappa=0$ modes in our previous
work\cite{kouchan-string-2}, as in the case of the $\kappa\neq 0$ modes. 
The differences in this case are the same as those in the case of
$\kappa\neq 0$ modes, i.e., those regarding the absence of the
curvature terms and the meaning of the covariant derivative
$D_{\alpha}$.

%*********************************************\\

On the basis of the Einstein equations
(\ref{eq:kneq-mode-equations-1})--(\ref{eq:kneq-mode-equations-7}) for
$\kappa\neq 0$ modes and equations (\ref{eq:14.171})--(\ref{eq:14.177}),
we compare the oscillatory behavior of a gravitating string and a test
string.

%*********************************************\\

%%%%%%%%%%%%%%%%%%%%%%%%%%%%%%%%%%%%%%%%%%%%%%%%%%%%%%%%%%%%%%%%%%%%%%
\section{Comparison with curved spacetime background analyses} 
\label{sec:Comparison}
%%%%%%%%%%%%%%%%%%%%%%%%%%%%%%%%%%%%%%%%%%%%%%%%%%%%%%%%%%%%%%%%%%%%%%

%*********************************************\\

Here, we compare the results of the present work, the Einstein
equations
(\ref{eq:kneq-mode-equations-1})--(\ref{eq:kneq-mode-equations-7}) and
(\ref{eq:14.171})--(\ref{eq:14.177}) derived in the previous
sections (\S\S\ref{sec:kappaneq0-mode-perturbation-variables} and
\ref{sec:kappa=0-mode-perturbation-variables}), with the results of 
our previous works\cite{kouchan-string-1,kouchan-string-2}.
Since the analyses to derive these equations are based on the
harmonic expansion, as in the last section, we again treat the
dynamical modes that propagate along the string (the $\kappa=0$ 
modes) and others (the $\kappa\neq 0$ modes), separately.
Of course, as the remaining perturbative modes, the cylindrical
and stationary perturbative modes exist as seen in
\S\ref{sec:lambda-order-solution}.  
However, these stationary perturbations have nothing to do with
the dynamics of the Nambu-Goto membrane.
For this reason we discuss only the $\kappa\neq 0$ and $\kappa=0$
modes in this section.

%*********************************************\\

%%%%%%%%%%%%%%%%%%%%%%%%%%%%%%%%%%%%%%%%%%%%%%%%%%%%%%%%%%%%%%%%%%%%%%
\subsection{$\kappa\neq 0$ mode solutions}
\label{sec:kappaneq0-mode-perturbation}

%*********************************************\\

For $\kappa\neq 0$ mode perturbations, the $O(\epsilon\lambda)$
Einstein equations are given by 
(\ref{eq:kneq-mode-equations-1})--(\ref{eq:kneq-mode-equations-7}).
These equations are almost same as the linearized Einstein
equations for the $\kappa\neq0$ modes in Ref.~1),
as we previously noted. 
The displacement perturbation $V^{\alpha}$ in
(\ref{eq:kneq-mode-equations-2}) and 
(\ref{eq:kneq-mode-equations-3}) seems induce nontrivial
perturbations of the metric.
However, unlike in our previous work for $\kappa\neq 0$ mode
perturbations, these equations lead to $V^{\alpha}=0$ as we show in
this section. 
Hence, we conclude that the $\kappa\neq 0$ mode metric
perturbations have nothing to do with the string dynamics at
$O(\epsilon\lambda)$.
To show this, we first derive the solutions to the equations
(\ref{eq:kneq-mode-equations-1})--(\ref{eq:kneq-mode-equations-7}).

%*********************************************\\

First, we consider Eqs.~(\ref{eq:kneq-mode-equations-6}) and
(\ref{eq:kneq-mode-equations-7}) for the variable $F_{\alpha}$. 
We consider the decomposition of the vector on ${\cal M}_{2}$ as 
\begin{equation}
  F_{\alpha} = D_{\alpha}\Psi_{(o)} 
  + {\cal E}_{\alpha\beta}D^{\beta}\bar{\Phi}_{(o)}.
\end{equation}
Using (\ref{eq:kneq-mode-equations-7}), we obtain
\begin{equation}
  D^{\alpha}F_{\alpha} = \Delta\Psi_{(o)} = 0.
\end{equation}
This shows that $\Psi_{(o)}$ is a solution to the two-dimensional
Laplace equation. 
Then, we can easily check that there exists a function 
${\cal I}$ such that
\begin{equation}
  \epsilon_{\alpha\beta}D^{\beta} {\cal I} = D_{\alpha}\Psi_{(o)},
\end{equation}
and we obtain
\begin{equation}
  F_{\alpha} = \epsilon_{\alpha\beta}D^{\beta}
  \left(\bar{\Phi}_{(o)} + {\cal I}\right)
  =: {\cal E}_{\alpha\beta}D^{\beta}\hat{\Phi}_{(o)}.
\end{equation}
Substituting this into (\ref{eq:kneq-mode-equations-6}), we obtain 
\begin{eqnarray}
  0 &=& (\Delta + \kappa^{2}) 
  {\cal E}_{\alpha\beta}D^{\beta}\left(\bar{\Phi}_{(o)} + {\cal I}\right)
  \nonumber\\
  &=& {\cal E}_{\alpha\beta}D^{\beta} (\Delta + \kappa^{2}) 
  \left(\bar{\Phi}_{(o)} + {\cal I}\right).
\end{eqnarray}
Integrating this equation, we obtain
\begin{equation}
  (\Delta + \kappa^{2}) \left(\bar{\Phi}_{(o)} + {\cal I}\right) = C,
\end{equation}
where $C$ is a constant.
This equation can also be written by
\begin{equation}
  (\Delta + \kappa^{2}) \left( \bar{\Phi}_{(o)} + {\cal I}
    - \frac{1}{\kappa^{2}} C \right) = 0.
\end{equation}
Here, we redefine the variable $\Phi_{(o)}$ as
\begin{equation}
  \Phi_{(o)} := \bar{\Phi}_{(o)} + {\cal I} - \frac{1}{\kappa^{2}} C.
\end{equation}
Then, Eqs.~(\ref{eq:kneq-mode-equations-6}) and
(\ref{eq:kneq-mode-equations-7}) are reduced to 
\begin{equation}
  F_{\alpha} = {\cal E}_{\alpha\beta}D^{\beta}\Phi_{(o)}, \quad
  (\Delta + \kappa^{2}) \Phi_{(o)} = 0.
\end{equation}
The scalar variable $\Phi_{(o)}$ represents the gravitational wave
that corresponds to the odd mode gravitational wave in
Ref.~1). 
These gravitational waves propagate freely and have nothing to do with
the motion of the string.

%*********************************************\\

Next, we consider Eqs.~(\ref{eq:kneq-mode-equations-1})--(\ref{eq:kneq-mode-equations-5}).
First, (\ref{eq:kneq-mode-equations-4}) shows that the tensor
$F_{\alpha\beta}$ is traceless.
Therefore we can decompose $F_{\alpha\beta}$ as
\begin{equation}
  F_{\alpha\beta} = \left(D_{\alpha}D_{\beta} -
    \frac{1}{2}\gamma_{\alpha\beta} \Delta\right) \bar{\Phi}_{(e)} 
  + {\cal A}_{\alpha\beta} \Psi_{(e)}.
\end{equation}
Substituting this into (\ref{eq:kneq-mode-equations-3}), we
obtain 
\begin{eqnarray}
  \label{eq:D.130}
  V_{\alpha} = \frac{1}{2} D_{\alpha} \Delta \bar{\Phi}_{(e)} 
  + \frac{1}{2} {\cal E}_{\alpha\beta}D^{\beta} \Delta \Psi_{(e)} 
  - \frac{1}{2} D_{\alpha} F. 
\end{eqnarray}
Further, Eqs.~(\ref{eq:kneq-mode-equations-2}) and (\ref{eq:D.130})
yield 
\begin{equation}
  \label{eq:14.205-barPhi-Psi}
  0 = \left(D_{\alpha}D_{\beta} -
    \frac{1}{2}\gamma_{\alpha\beta}\Delta\right) 
  \left(\bar{\Phi}_{(e)}  + \frac{1}{\kappa^{2}} F \right) 
  + {\cal A}_{\alpha\beta} \Psi_{(e)}. 
\end{equation}
Taking the divergence and then the rotation of this
equation, we obtain 
\begin{equation}
  \label{eq:deltadeltaPsie}
  \Delta\Delta\Psi_{(e)} = 0.
\end{equation}
Since $\Delta$ is the Laplacian associated with the flat metric,
we easily show that for any solution to
(\ref{eq:deltadeltaPsie}), there exists a function ${\cal I}$
such that
\begin{equation}
  \label{eq:there-exits-calI}
  {\cal A}_{\alpha\beta} \Psi_{(e)} = \left(D_{\alpha}D_{\beta} 
    - \frac{1}{2}\gamma_{\alpha\beta} \Delta \right) {\cal I}.
\end{equation}
This implies that we may choose $\Psi_{(e)}=0$ without loss of
generality, $F_{\alpha\beta}$ is given by 
\begin{equation}
  \label{eq:K-ab-decomp-reduced}
  F_{\alpha\beta} = \left(D_{\alpha}D_{\beta} 
    - \frac{1}{2}\gamma_{\alpha\beta} \Delta \right) 
  \left(\bar{\Phi}_{(e)} + {\cal I}\right), 
\end{equation}
and Eq.~(\ref{eq:14.205-barPhi-Psi}) is reduced to 
\begin{equation}
  \label{eq:14.205-barPhi-K}
  \left(D_{\alpha}D_{\beta} - \frac{1}{2}\gamma_{\alpha\beta} \Delta \right) 
  \left(\bar{\Phi}_{(e)} + {\cal I} + \frac{1}{\kappa^{2}} F \right) = 0.
\end{equation}
The divergence of Eq.~(\ref{eq:14.205-barPhi-K}) is given by 
\begin{equation}
  D_{\alpha}\Delta \left(\bar{\Phi}_{(e)} + {\cal I} 
    + \frac{1}{\kappa^{2}} F \right) = 0.
\end{equation}
This can be easily integrated as 
\begin{eqnarray}
  \label{eq:F-barPhi-C-rela}
  C = \Delta \left(\bar{\Phi}_{(e)} + {\cal I}\right) - F,
\end{eqnarray}
where $C$ is a constant of integration and we have used
Eq.~(\ref{eq:kneq-mode-equations-1}). 
Substituting this into Eq.~(\ref{eq:14.205-barPhi-K}), we obtain 
\begin{equation}
  \label{eq:D.137}
  \left(D_{\alpha}D_{\beta} - \frac{1}{2}\gamma_{\alpha\beta} \Delta \right) 
  (\Delta + \kappa^{2}) \left(\bar{\Phi}_{(e)} + {\cal I}\right)= 0.
\end{equation}
On the other hand, substituting Eq.~(\ref{eq:F-barPhi-C-rela}) into 
Eq.~(\ref{eq:kneq-mode-equations-1}), we obtain
\begin{eqnarray}
  \label{eq:D.138}
  \Delta (\Delta + \kappa^{2}) 
  \left(\bar{\Phi}_{(e)} + {\cal I}\right)= \kappa^{2} C.
\end{eqnarray}
From Eqs.~(\ref{eq:D.137}) and (\ref{eq:D.138}), we obtain 
\begin{equation}
  \label{eq:barPhi-C-rela}
  D_{\alpha}D_{\beta} (\Delta + \kappa^{2}) 
  \left(\bar{\Phi}_{(e)} + + {\cal I}\right)
  = \frac{\kappa^{2} C}{2} \gamma_{\alpha\beta}.
\end{equation}
Using a function ${\cal G}$ that satisfies the equation
\begin{equation}
  \label{eq:exist-cal-G-function-eq}
  D_{\alpha}D_{\beta} {\cal G} = \frac{1}{2} \gamma_{\alpha\beta}C,
\end{equation}
Eqs.~(\ref{eq:F-barPhi-C-rela}) and (\ref{eq:barPhi-C-rela}) can be
written as  
\begin{eqnarray}
  \label{eq:master-equation-pre-kneq0}
  && D_{\alpha}D_{\beta} (\Delta + \kappa^{2}) (\bar{\Phi}_{(e)} + {\cal I}
  - {\cal G}) = 0, \\
  && F = \Delta ( \bar{\Phi}_{(e)} + {\cal I} - {\cal G} ).
\end{eqnarray}
Here, we note that the existence of a solution to
Eq.~(\ref{eq:exist-cal-G-function-eq}) is confirmed by explicit 
calculation, introducing an explicit coordinate system on 
$({\cal M}_{2},\gamma_{ab})$.
Further, Eq.~(\ref{eq:K-ab-decomp-reduced}) can be written
\begin{equation}
   F_{\alpha\beta}
   = \left(D_{\alpha}D_{\beta}-\frac{1}{2}\gamma_{\alpha\beta}\Delta\right)
   (\bar{\Phi}_{(e)} + {\cal I} - {\cal G})
\end{equation}
using Eq.~(\ref{eq:exist-cal-G-function-eq}).

%*********************************************\\

Integrating Eq.~(\ref{eq:master-equation-pre-kneq0}), we
obtain 
\begin{equation}
  (\Delta + \kappa^{2}) (\bar{\Phi}_{(e)} + {\cal I} - {\cal G}) 
  = E_{\alpha} X^{\alpha},
\end{equation}
where $E_{\alpha}$ represents the component of a constant vector, and
the vector $X^{\alpha}$ satisfies 
$D_{\alpha}X^{\beta} = \delta_{\alpha}^{\;\;\beta}$.
Then, defining the function $\Phi_{(e)}$ by 
\begin{equation}
  \Phi_{(e)} := \bar{\Phi}_{(e)} + {\cal I} - {\cal G} - \frac{1}{\kappa^{2}}
  E_{\beta} X^{\beta}, 
\end{equation}
Eqs.~(\ref{eq:kneq-mode-equations-1}),
(\ref{eq:kneq-mode-equations-2}) and 
(\ref{eq:kneq-mode-equations-3}) are reduced to the single equation
\begin{equation}
  \label{eq:kneq0-even-master-eq}
  (\Delta + \kappa^{2}) \Phi_{(e)} = 0,
\end{equation}
and the gauge invariant metric perturbation variables $F$ and
$F_{\alpha\beta}$ are given by 
\begin{equation}
  \label{eq:F-Fab-Phi-rela-final}
  F = \Delta \Phi_{(e)}, \quad 
  F_{\alpha\beta} = 
  \left(D_{\alpha}D_{\beta}-\frac{1}{2}\gamma_{\alpha\beta}\Delta\right) 
  \Phi_{(e)},
\end{equation}
respectively.
This scalar variable $\Phi_{(e)}$ represents gravitational waves
correspond to the even mode gravitational waves in
Ref.~1).
Thus, using the solution to Eq.~(\ref{eq:kneq0-even-master-eq}), we
obtain the $\kappa\neq 0$ mode solution of $O(\epsilon\lambda)$
perturbations.

%*********************************************\\

Substituting Eq.~(\ref{eq:F-Fab-Phi-rela-final})
into Eqs.~(\ref{eq:kneq-mode-equations-3}) and
(\ref{eq:kneq-mode-equations-5}), we directly see that  
\begin{equation}
  V^{a} = 0 = \Sigma.
\end{equation}
Thus, the string does not bend in the $\kappa\neq 0$ mode
perturbations at $O(\epsilon\lambda)$.
This conclusion is different from that in our previous works, but
it can be inferred from them\cite{kouchan-string-1}.
Actually, the conclusion obtained here can also be obtained by
taking the limit in which the background deficit angle
vanishes. 
Hence, we find that $\kappa\neq 0$ mode gravitational waves do not
couple to the string motion, at least at $O(\epsilon\lambda)$.
These gravitational waves have nothing to do with the string
oscillation in this perturbative treatment. 
This conclusion also implies that at least at first order in
the string oscillation amplitude, the string oscillations do not
produce gravitational waves that propagate to regions far from
the string.

%*********************************************\\

Finally, we note that the solutions to
Eqs.~(\ref{eq:kneq-mode-equations-1})--(\ref{eq:kneq-mode-equations-7})
describe the free propagation of gravitational waves.
These gravitational wave solutions also exist at both 
$O(\epsilon)$ and $O(\lambda)$.
At both $O(\epsilon)$ and $O(\lambda)$, these gravitational waves also
have nothing to do with the oscillations of the string. 
For this reason, we have imposed the condition that such gravitational
waves exist at neither $O(\epsilon)$ nor $O(\lambda)$.

%*********************************************\\

%%%%%%%%%%%%%%%%%%%%%%%%%%%%%%%%%%%%%%%%%%%%%%%%%%%%%%%%%%%%%%%%%%%%%%
\subsection{$\kappa=0$ mode solutions}
\label{sec:kappa=0-mode-perturbation}

%*********************************************\\

For $\kappa=0$ mode perturbations, the $O(\epsilon\lambda)$ Einstein
equations are given by (\ref{eq:14.171})--(\ref{eq:14.177}). 
These equations are also almost the same as the linearized Einstein
equations for the $\kappa=0$ modes in Ref.~2) as we noted.
However, in this subsection, we show that
Eqs.~(\ref{eq:14.171})--(\ref{eq:14.177}) include solutions describing
the string oscillation without gravitational waves.
These oscillatory solutions correspond to the pp-wave exact solutions 
on the spacetime with deficit angle, as in our previous paper.
These oscillatory solutions also coincide with the oscillations
of the test string.
Hence, the absence of the dynamical degree of freedom of
gravitating string oscillations is not contrary to the
dynamics of the test string, at least in the infinite string case. 
Further, from the derivation of the solutions, we can easily see that
there are no other oscillatory solutions that describes the
oscillations of an infinite string. 
To show this, we only need to derive the solutions to
Eqs.~(\ref{eq:14.171})-(\ref{eq:14.177}).

%*********************************************\\

In this paper, we derive the solutions only on the support of
$V_{\alpha}$ and $\Sigma$ (i.e., the support of $\rho$) and in the
vacuum region (i.e., outside of the support of $\rho$), respectively.
These solutions correspond to the solutions describing the string oscillations
with gravitational wave propagation in Ref.~2).
Strictly speaking, to show the existence of a solution describing
string oscillations without gravitational waves, we should construct 
global perturbative solutions on ${\cal M}$ by matching these
solutions at the surface of the support of $\rho$. 
As shown in Ref.~2), this can be accomplished
using the Israel junction conditions\cite{Israel}.
On the other hand, the perturbed Einstein equations
(\ref{eq:14.171})--(\ref{eq:14.177}) and those in
Ref.~2) are both apply to the case of an arbitrary
distribution of $\rho$, and their are nearly identical.
Further, Israel's conditions guarantee only that the matching is not
contrary to the Einstein equations. 
Therefore, it is obvious that the matching conditions derived from the
$O(\epsilon\lambda)$ Israel's junction conditions also have the 
same as those in Ref.~2).
Hence, to clarify the string oscillations without gravitational waves
and compare with the results in Ref.~2), it is
enough to derive the solutions in the two regions, i.e., on the support of
$\rho$ and outside of this support.
From these solutions, we can easily see that the string can oscillate
without gravitational waves in the case of a flat spacetime background.

%*********************************************\\

First, we consider the solutions to Eqs.~(\ref{eq:14.174}) 
and (\ref{eq:14.176}).
Since we impose the condition that the metric perturbation $H$ and
$H_{\alpha}$ are regular on the space ${\cal M}_{2}$, we obtain the
solution to these equations
\begin{equation}
  \label{eq:H_alpha=0=H}
  H_{\alpha} = 0 = H.
\end{equation}
Then, Eq.~(\ref{eq:14.172}) becomes trivial.
Since $H_{\alpha\beta}$ is traceless, as expressed by
Eq.~(\ref{eq:14.175}), we decompose the matrix $H_{\alpha\beta}$ as 
\begin{equation}
  \label{eq:14.decomp-of-Kab}
  H_{\alpha\beta} = \left(D_{\alpha}D_{\beta} -
    \frac{1}{2}\gamma_{\alpha\beta}\Delta \right)\Phi_{(\kappa 0)} 
  + {\cal A}_{\alpha\beta}\Psi_{(\kappa 0)} .
\end{equation}
Substituting Eqs.~(\ref{eq:14.decomp-of-Kab}) and
(\ref{eq:H_alpha=0=H}) into Eq.~(\ref{eq:14.173}), we obtain  
\begin{eqnarray}
  \label{eq:Valpha-Phi-Psi-rela}
  V_{\alpha} = \frac{1}{2}D_{\alpha} \Delta \Phi_{(\kappa 0)}  
  + \frac{1}{2} {\cal E}_{\beta\alpha} D^{\beta} \Delta
  \Psi_{(\kappa 0)} .
\end{eqnarray}
This expression of $V_{\alpha}$ shows that $\Phi_{(\kappa 0)}$
and $\Psi_{(\kappa 0)}$ correspond to the irrotational and
rotational parts of the matter velocity, respectively.
Further, using Eqs.~(\ref{eq:14.decomp-of-Kab}) and
(\ref{eq:Valpha-Phi-Psi-rela}), we easily see that
Eq.~(\ref{eq:14.171}) is trivial.
Finally, Eq.~(\ref{eq:14.177}) gives an expression
of $\Sigma$ in terms $\Phi_{(\kappa 0)}$:
\begin{equation}
  \label{eq:Sigma-Phi-rela}
  \Sigma = - \Delta \Delta \Phi_{(\kappa 0)}. 
\end{equation}
Equations (\ref{eq:14.decomp-of-Kab}), (\ref{eq:Valpha-Phi-Psi-rela}),
and (\ref{eq:Sigma-Phi-rela}) are solutions to Eqs.(\ref{eq:14.174}) 
and (\ref{eq:14.176}) and have forms similar to the corresponding
equations in Ref.~2).

%*********************************************\\

When the displacement $V_{\alpha}$ vanishes, i.e., outside of the
support of $\rho$, (\ref{eq:Valpha-Phi-Psi-rela}) yields
$\Delta\Delta\Psi_{(\kappa 0)}=0$.
This implies that we may choose $\Psi_{(\kappa 0)}=0$ without loss of
generality, as in the case of $\kappa\neq 0$ mode perturbations [see
Eq.~(\ref{eq:there-exits-calI})].
Further, Eq.~(\ref{eq:Valpha-Phi-Psi-rela}) also yields
$\Delta\Phi_{(\kappa 0)}=0$, i.e.,
\begin{equation}
  \label{eq:DeltaPhi-k0-vac}
  \Phi_{(\kappa 0)} = A_{0}\ln r 
  + \sum_{m=1}^{\infty}\left(A_{m}r^{-m} + B_{m}r^{m}\right)e^{im\phi},
\end{equation}
where $r:=\sqrt{x^{2}+y^{2}}$ and $\phi:=\arctan(y/x)$ are the usual
radial and azimuthal coordinates on the two-dimensional flat space,
respectively [see Eq.(\ref{eq:greek-indices-def})].
Then, Eq.~(\ref{eq:14.decomp-of-Kab}) reduces to 
\begin{equation}
  \label{eq:14.decomp-of-Kab-vac}
  H_{\alpha\beta} = D_{\alpha}D_{\beta}\Phi_{(\kappa 0)}.
\end{equation}

%*********************************************\\

From the above, we see that in the derivation of the solutions
describing by
Eqs.~(\ref{eq:14.decomp-of-Kab})--(\ref{eq:Sigma-Phi-rela}), 
(\ref{eq:DeltaPhi-k0-vac}) and (\ref{eq:14.decomp-of-Kab-vac}),
there is no delicate problem due to the fact that we used a
background spacetime that is different from those in
Ref.~2).
To study the global solutions for $\kappa=0$ mode perturbations, we
may consider the expressions of the solutions in
Ref.~2) with a vanishing background deficit 
angle (or a vanishing background curvature).
In particular, in the thin string situation, only the $m=1$ mode in
Eq.~(\ref{eq:DeltaPhi-k0-vac}) (i.e., the dipole deformation) is
relevant to the string displacement. 
Unlike in the curved background case, the solution
(\ref{eq:14.decomp-of-Kab-vac}) for the $m=1$ mode vanishes in this
case, due to the absence of the background curvature. 
On the other hand, the displacement of the thin string defined in
Ref.~2) gives a finite contribution and it does
not depend on the thickness of the string, unlike in the curved
background case. 
This implies that we can define the displacement of the thin string at
any point on the support of the $O(\lambda)$ string energy density
$\rho$, although in previous works we defined it on the boundary
surface of the support of $\rho$.
Then, we can derive the solutions describing the string oscillations
without gravitational waves in the flat spacetime background.
These solutions merely represent the oscillations of a test string.
Hence, we can conclude that the absence of a dynamical degree of
freedom of gravitating string oscillations is not contradict to the
dynamics of a test string, at least in the infinite string case.

%*********************************************\\

We note that $\Phi_{(\kappa 0)}$ in (\ref{eq:14.decomp-of-Kab-vac})
corresponds to the gravitational waves that propagate along the string. 
Further, we can easily see that this solution is just an
exact solution to the Einstein equation that is called a
``cosmic string traveling waves.''\cite{Garfinkle} 
The cosmic string traveling wave is the pp-wave exact solution.
The test string oscillations obtained above represent just one limit
of this pp-wave solution.

%*********************************************\\

Finally, the solution (\ref{eq:14.decomp-of-Kab-vac}) with
(\ref{eq:DeltaPhi-k0-vac}) in the case $V^{\alpha}=0=\Sigma$ describes
the propagation of free gravitational waves along ${\cal M}_{1}$.
When $V^{\alpha}=0=\Sigma$ at any point on ${\cal M}_{2}$, we impose
the regularity condition at $r=0$ and $r\rightarrow\infty$ on the
gauge invariant metric perturbation $H_{\alpha\beta}$.
With this condition, only the $m=2$ mode gives a constant contribution 
to $H_{\alpha\beta}$.
This is just plane wave propagation of gravitational wave along 
${\cal M}_{1}$. 
These gravitational wave solutions also exist at $O(\epsilon)$
and $O(\lambda)$.
However, these gravitational waves have nothing to do with the
oscillations of the string.
We thus see that at $O(\epsilon)$ and $O(\lambda)$, we have imposed
the condition such that such gravitational waves do not exist.

%*********************************************\\

%%%%%%%%%%%%%%%%%%%%%%%%%%%%%%%%%%%%%%%%%%%%%%%%%%%%%%%%%%%%%%%%%%%%%%
\section{Summary and discussion}
\label{sec:summary-discussion}
%%%%%%%%%%%%%%%%%%%%%%%%%%%%%%%%%%%%%%%%%%%%%%%%%%%%%%%%%%%%%%%%%%%%%%

%*********************************************\\

In this article, we have considered the perturbative
oscillations of an infinite Nambu-Goto string.
We developed the two parameter gauge invariant perturbation
technique on a flat spacetime background.
This perturbation theory includes two infinitesimal perturbation
parameters, the string oscillation amplitude $\epsilon$ and the string
energy density $\lambda$. 
We have considered the Einstein equations of $O(\epsilon)$,
$O(\lambda)$, and $O(\epsilon\lambda)$.

%*********************************************\\

In spite of the difference of the background spacetime,
the Einstein equations of $O(\epsilon\lambda)$ are almost the same as
the linearized Einstein equations in our previous
papers\cite{kouchan-string-1,kouchan-string-2}.
From these equations, we can obtain the solutions those describe
the string oscillations without gravitational waves.
From this fact, we conclude that the $m=1$ mode cosmic string
traveling wave should be regarded as the string oscillation, as
discussed by Vachaspati and Garfinkle\cite{Garfinkle}. 
This solution coincides with that in our previous
works\cite{kouchan-string-2} in the artificial limit in which
the background deficit angle vanishes. 
Further, we have seen that there is no other oscillatory solutions
that describes the oscillations of an infinite string.
Since the existence of the pp-wave solution is closely related to the
specific symmetry of the spacetime, it is also natural to conjecture
that such solutions do not exist in more generic situations.

%*********************************************\\

We should emphasize that the existence of the solution describing the
string oscillations without gravitational waves is due to the fact
that we have chosen a Minkowski spacetime as a background for the
perturbation. 
This solution corresponds to that describing the thin string
oscillation with the propagation of gravitational waves along the
string in Ref.~2). 
Further, the results obtained here are less accurate than those 
obtained in Ref.~2), because the results in
Ref.~2) can only be obtained from the infinite 
sum of $O(\epsilon\lambda^{n})$ terms in the perturbative treatment 
developed here.
Therefore, the conclusion in Ref.~2) is more
accurate picture than that based on the dynamics of a test string,
i.e., we conclude that 
{\it the dynamical degree of freedom of string oscillations is
  that of gravitational waves},
and 
{\it an infinite string can oscillate, but these oscillations simply
  represent the propagation of the gravitational waves along the string}, 
which corresponds to cosmic string traveling waves (the exact
solution). 
We have seen that these conclusions do not contradict to the results
based on analyses developed here.

%*********************************************\\

Further, we point out that the equation of motion of $O(\epsilon)$ is
given by 
\begin{eqnarray}
  \frac{\partial}{\partial\epsilon} K^{a} 
  = - {\cal D}^{b}{\cal D}_{b} \widehat{V}^{a} = 0.
  \label{eq:test-string-eq-of-motion}
\end{eqnarray}
This is identically the equation of motion for a test string to first
order in the string oscillation amplitude. 
The solutions to this equation are consistent with the solutions
to the Einstein equation of $O(\epsilon\lambda)$.
Hence, the test string dynamics of $O(\epsilon)$ are consistent with
the $O(\epsilon\lambda)$ Einstein equations. 
Further, the oscillations of the string, which are the solutions to
Eq.~(\ref{eq:test-string-eq-of-motion}) do not produce gravitational
waves that propagate to regions far from the string, at least at
$O(\epsilon\lambda)$.

%*********************************************\\

We also note that this result is consistent with the power of the
gravitational wave emitted from an infinite Nambu-Goto string 
derived by Sakellariadou\cite{Sakellariadou}, which is based on
the energy momentum tensor of an oscillating test string.
She derived the power of the gravitational waves emitted by the
helicoidal standing oscillations of an infinite string.
In her result, the second order term of the string oscillation
amplitude does not appear in the power of the gravitational waves. 
The second order [$O(\epsilon^{2})$] term in the power of
gravitational waves corresponds to the first order [$O(\epsilon)$] of
the string oscillation amplitude. 
From her analyses, it is not clear if the absence of an
$O(\epsilon^{2})$ term in the power of gravitational waves is due to
the helicoidal standing oscillations of the string, which is a special
solution to the equation of motion of the Nambu-Goto string. 
However, our result shows that the absence of an $O(\epsilon^{2})$
term in the power of gravitational waves is generic for gravitational
waves emitted from an oscillating infinite string.
Our result implies that string oscillation of $O(\epsilon)$ does not
produce gravitational waves that reach regions far from the string,
even if the string oscillations are very complicated.

%*********************************************\\

We should stress that the consistency of the dynamics of
a gravitationing string and that of a test string discussed here is
due to the existence of the cosmic string traveling wave solution. 
We note that the model studied here includes traveling waves, and the
string can continue to oscillate, due to the existence of the pp-wave
exact solution (cosmic string traveling wave). 
We also note that the existence of traveling waves has not been
confirmed in the other exactly soluble models of gravitating
Nambu-Goto walls\cite{Kodama,Ishibashi-Tanaka}.
In this sense, the model considered here might be an exceptional
case. 
If there do not exist traveling waves in more generic situations
of gravitating Nambu-Goto membranes, there is no guarantee
that the oscillatory behavior of gravitating Nambu-Goto membranes is
approximated by that of a test membrane. 
Further, we have found that there is no dynamical degree of
freedom of the string oscillations, except for the cosmic string
traveling waves.
Hence, we can conclude that if the existence of traveling waves is
not confirmed, the oscillatory behavior of gravitating
Nambu-Goto membranes becomes highly nontrivial, and there is no
guarantee that the estimate of the power of the gravitational waves
emitted from extended objects based on the test membrane dynamics is
valid.
Of course, it may result that these two treatments are consistent as
seen here.
Therefore, it would be interesting to consider the other models in
which the perturbation theory on Minkowski spacetime is well-defined
and for which the existence of traveling wave solutions along the
membrane has not yet been confirmed. 
We leave this topic for future investigations.

%*********************************************\\

%%%%%%%%%%%%%%%%%%%%%%%%%%%%%%%%%%%%%%%%%%%%%%%%%%%%%%%%%%%%%%%%%%%%%%
\section*{Acknowledgements}
%%%%%%%%%%%%%%%%%%%%%%%%%%%%%%%%%%%%%%%%%%%%%%%%%%%%%%%%%%%%%%%%%%%%%%
The author would like to thank Professor Minoru Omote (Keio
University) for his continuous encouragements and anonymous referees
for suggesting changes that have improved the manuscript.

%%%%%%%%%%%%%%%%%%%%%%%%%%%%%%%%%%%%%%%%%%%%%%%%%%%%%%%%%%%%%%%%%%%%%%
\appendix

%%%%%%%%%%%%%%%%%%%%%%%%%%%%%%%%%%%%%%%%%%%%%%%%%%%%%%%%%%%%%%%%%%%%%%
\section{A More Precise Description of ``Thick String''}
\label{sec:Appendix-thick}
%%%%%%%%%%%%%%%%%%%%%%%%%%%%%%%%%%%%%%%%%%%%%%%%%%%%%%%%%%%%%%%%%%%%%%

%*********************************************\\

Here, we give a more precise description of the ``thick string'' and
its intrinsic metric $q_{ab}$ studied in the main text. 
Since the world sheet of the usual Nambu-Goto string is a
two-dimensional timelike surface embedded in the
four-dimensional spacetime, we first consider the decomposition
of the spacetime and its tangent space into a lower dimensional
surface. 
Next, we consider the energy-momentum tensor for a ``thick''
Nambu-Goto membrane and its equation of motion derived from the
divergence of this tensor.
In this paper, the thickness of membranes is introduced in such a
manner that the equation of motion of membranes is maintained. 
Through this introduction of the thickness of the membrane, we obtain
the energy-momentum tensor for the ``regularized Nambu-Goto membrane.''
In this appendix, we consider the embedding of the $M$-dimensional
surface $\Sigma_{M}$ into the $N+M$-dimensional spacetime 
$({\cal M},g_{ab})$ to guarantee the extension to any dimensional
hypersurface embedded in a higher-dimensional spacetime.

%*********************************************\\

%%%%%%%%%%%%%%%%%%%%%%%%%%%%%%%%%%%%%%%%%%%%%%%%%%%%%%%%%%%%%%%%%%%%%
\subsection{Decomposition of the spacetime and metric}
\label{sec:Appendix-thick-1}

%*********************************************\\

Let $x:{\cal M}\rightarrow{\cal R}^{m}$ be a coordinate system on 
${\cal M}$.
We introduce $M$ functions $\sigma^{i}$ $(i=0,\cdots,M-1)$, so that
the hypersurface $x^{\mu} = x^{\mu}(\sigma^{i})$ 
($\mu=0,\cdots,N+M-1$) is $\Sigma_{M}$.
We also consider $N$ functions $\xi^{\alpha}$ 
$(\alpha = M,\cdots,M+N-1)$ such that all $\xi^{\alpha}$ are constant 
on $\Sigma_{M}$ and all the one-forms
$(d\xi^{\alpha})_{a}:=\nabla_{a}\xi^{\alpha}$ are linearly
independent of each other.
Further, choosing the functions $\sigma^{i}$ $(i=0,\cdots,M-1)$ so
that the one-forms $(d\sigma^{i})_{a}:=\nabla_{a}\sigma^{i}$ and
$(d\xi^{\alpha})_{a}$ are independent of each other.
Then, the set of functions $\{\sigma^{i},\xi^{\alpha}\}$ is a
coordinate system on ${\cal M}$ (at least in the neighborhood of
$\Sigma_{M}$).  
Here, we consider the tangent space spanned by $(d\xi^{\alpha})_{a}$
and introduce the orthonormal basis $(\theta^{\alpha})_{a}$ of this
tangent space by constructed from linear combinations of
$(d\xi^{\alpha})_{a}$; i.e., 
$g^{ab}(\theta^{\alpha})_{a}(\theta^{\beta})_{b} = \delta^{\alpha\beta}$,
where $\delta^{\alpha\beta}$ is the $N$-dimensional Kronecker delta.

%*********************************************\\

The metric induced on $\Sigma_{M}$ from the metric $g_{ab}$ on
${\cal M}$ is defined by 
\begin{equation}
  \label{eq:q-defE5}
  q_{ab} := g_{ab} -
  \delta_{\alpha\beta} (\theta^{\alpha})_{a} (\theta^{\beta})_{b}.  
\end{equation}
We also define the tensor 
\begin{equation}
  \label{gamma-def-E9}
  \gamma_{ab} := \delta_{\alpha\beta} (\theta^{\alpha})_{a} 
  (\theta^{\beta})_{b} =  g_{ab} - q_{ab}.
\end{equation}
We note that both $q_{a}^{\;\;b}:=q_{ac}g^{cb}$ and 
$\gamma_{a}^{\;\;b}:=\gamma_{ac}g^{cb}$ have the properties of the
projection operators: 
\begin{eqnarray}
  q_{a}^{\;\;b}q_{b}^{\;\;c} = q_{a}^{\;\;c}, \quad
  \gamma_{a}^{\;\;b}\gamma_{b}^{\;\;c} = \gamma_{a}^{\;\;c}, \quad
  \gamma_{a}^{\;\;b}q_{b}^{\;\;c} = 0.
\end{eqnarray}
The operator $q_{a}^{\;\;b}$ projects the vectors in ${\cal M}$
into $\Sigma_{M}$, and the operator $\gamma_{a}^{\;\;b}$ projects
to the complement space of $\Sigma_{M}$.
In the coordinate system $\{\sigma^{i},\xi^{\alpha}\}$, the spacetime
metric $g_{ab}$ and its inverse $g^{ab}$ are decomposed as
\begin{eqnarray}
  \label{eq:metric-components-N+M-decomp}
  g_{ab} &=& 
  \gamma_{\alpha\beta} 
  (d\xi^{\alpha})_{a} (d\xi^{\beta})_{b}
  + q_{ij}
  ( (d\sigma^{i}) + \beta^{i}_{\gamma} (d\xi^{\gamma}) )_{a}
  ( (d\sigma^{j}) + \beta^{j}_{\delta} (d\xi^{\delta}) )_{b}, \\
  \label{eq:E34}
  g^{ab} &=&
  \gamma^{\alpha\beta}
  \left( \frac{\partial}{\partial\xi^{\alpha}} -
    \beta^{i}_{\alpha} \frac{\partial}{\partial\sigma^{i}} \right)^{a}
  \left( \frac{\partial}{\partial\xi^{\beta}} -
    \beta^{i}_{\beta} \frac{\partial}{\partial\sigma^{i}} \right)^{b}
  + q^{ij} 
  \left(\frac{\partial}{\partial\sigma^{i}}\right)^{a}
  \left(\frac{\partial}{\partial\sigma^{j}}\right)^{b},
\end{eqnarray}
where 
\begin{equation}
  q_{ij} := q_{ab}
  \left(\frac{\partial}{\partial\sigma^{i}}\right)^{a}
  \left(\frac{\partial}{\partial\sigma^{j}}\right)^{b}, \quad
  \gamma^{\alpha\beta}:=g^{ab}(d\xi^{\alpha})_{a}(d\xi^{\beta})_{b}.
\end{equation}
The matrices $q^{ij}$ and $\gamma_{\alpha\beta}$ are the inverses
of $q_{ij}$ and $\gamma^{\alpha\beta}$, respectively. 
The coefficient $\beta^{i}_{\alpha}$ corresponds to the shift vector
in the ADM decomposition.

%*********************************************\\

For an arbitrary vector field $t^{a}$ that satisfies
$t^{a}=q_{b}^{\;\;a}t^{b}$, the covariant derivative 
${\cal D}_{a}$ associated with the metric $q_{ab}$ is defined by 
\begin{equation}
  {\cal D}_{a}t^{b} := q_{a}^{\;\; c} q_{d}^{\;\; b} \nabla_{c} t^{d}.
\end{equation}
Similarly, for an arbitrary vector field $n^{a}$ that satisfies 
$n^{a}=\gamma_{b}^{\;\;a}n^{b}$, the covariant derivative
$D_{a}$ associated with the metric $\gamma_{ab}$ is defined by 
\begin{equation}
  D_{a}n^{b} := \gamma_{a}^{\;\; c} \gamma_{d}^{\;\; b} \nabla_{c} n^{d}.
\end{equation}
We also note that these derivatives satisfy ${\cal D}_{a}q_{bc}=0$ and  
$D_{a}\gamma_{bc}=0$, respectively.
These definitions of the covariant derivatives are naturally
extended to tensor fields of arbitrary rank.

%*********************************************\\

Here, we introduce the second fundamental form by 
\begin{equation}
  \label{eq:K_abc-def}
  K_{a\;\;c}^{\;\;b} := q_{a}^{\;\;e}q_{d}^{\;\;b}\nabla_{e}q_{c}^{\;\;d}.
\end{equation}
Using this second fundamental form,  the relation between the
covariant derivative ${\cal D}_{a}$ and $\nabla_{a}$ is given by 
\begin{equation}
  \label{eq:gauss's-formula}
  X^{b}\nabla_{b}Y^{a} = X^{b}{\cal D}_{b}Y^{a} 
  + X^{c} Y^{d} K_{cd}^{\;\;\;\;a}
\end{equation}
for any vector field $X^{a}=q_{b}^{\;\;a}X^{b}$ and
$Y^{a}=q_{b}^{\;\;a}Y^{b}$, where the first term in the right-hand
side of Eq.~(\ref{eq:gauss's-formula}) is the component of
$X^{b}\nabla_{b}Y^{a}$ tangential to the tangent space of $\Sigma_{M}$,
and the second term, $X^{c}Y^{d}K_{cd}^{\;\;\;\;a}$, is the normal
component.

%*********************************************\\

It is instructive to consider the trace of this extrinsic
curvature in the coordinate system in which the spacetime metric
$g_{ab}$ is given by
\begin{equation}
  g_{ab} = \left\{ 
    \gamma_{\alpha\beta} \frac{\partial\xi^{\alpha}}{\partial x^{\mu}}
    \frac{\partial\xi^{\beta}}{\partial x^{\nu}}
    + q_{ij} \left(\frac{\partial\sigma^{i}}{\partial x^{\mu}} +
      \beta^{i}_{\gamma} \frac{\partial\xi^{\gamma}}{\partial x^{\mu}}\right)
    \left(\frac{\partial\sigma^{j}}{\partial x^{\nu}} +
      \beta^{j}_{\delta} \frac{\partial\xi^{\delta}}{\partial x^{\nu}}\right)
    \right\} (dx^{\mu})_{a} (dx^{\nu})_{b}.
\end{equation}
In this coordinate system, the extrinsic curvature is given by 
\begin{eqnarray}
  K_{ab}^{\;\;\;\;c} =: K_{\mu\nu}^{\;\;\;\;\;\;\rho} 
  (dx^{\mu})_{a} (dx^{\nu})_{b}
   \left(\frac{\partial}{\partial x^{\rho}}\right)^{c},
\end{eqnarray}
and its trace is given by 
\begin{eqnarray}
   K^{c} &:=& g^{ab} K_{ab}^{\;\;\;\;c} = q^{ab} K_{ab}^{\;\;\;\;c} 
   =: K^{\mu} \left(\frac{\partial}{\partial x^{\mu}}\right),
\end{eqnarray}
where $K^{\mu}$ is given by 
\begin{equation}
  K^{\mu} =   \frac{1}{\sqrt{-q}}
  \frac{\partial}{\partial\sigma^{i}}
  \left( \sqrt{-q}
    q^{ij} \frac{\partial x^{\mu}}{\partial\sigma^{j}}
  \right)
  + \Gamma_{\kappa\sigma}^{\mu} q^{ij}
  \frac{\partial x^{\kappa}}{\partial\sigma^{i}}
  \frac{\partial x^{\sigma}}{\partial\sigma^{j}}.
  \label{eq:trace-K^mu-x-chart-eq}
\end{equation}

%*********************************************\\

%%%%%%%%%%%%%%%%%%%%%%%%%%%%%%%%%%%%%%%%%%%%%%%%%%%%%%%%%%%%%%%%%%%%%
\subsection{Infinitesimally thin Nambu-Goto membrane}
\label{sec:Appendix-thick-2}

%*********************************************\\

Here, we consider the action of an infinitesimally thin
Nambu-Goto membrane.
The Nambu-Goto action is given by the area of the world sheet of
the membrane:
\begin{equation}
  \label{eq:NG-action-NG-mem-1}
  S = - \mu \int d^{M}\sigma\sqrt{-q}, \quad q = \det(q_{ij}).
\end{equation}
Here, $\mu$ is the energy density of the membrane.
The variation of this action with respect to $x^{\mu}(\sigma^{i})$
gives the equation of motion of the membrane: 
\begin{equation}
  \label{eq:NG-action-NG-mem-2}
  K^{a} = 0.
\end{equation}
The energy-momentum tensor of the membrane can be found by
varying the action (\ref{eq:NG-action-NG-mem-1}) with respect to
the metric $g_{ab}$.
We have
\begin{equation}
  \label{eq:NG-action-NG-mem-3}
  T^{ab}\sqrt{-g} = - 2 \frac{\delta S}{\delta g_{ab}} = 
  \mu \int d^{M}\sigma
  \sqrt{-q} q^{ab} \delta^{M+N}(x^{\alpha}-x^{\alpha}(\sigma^{i})),
\end{equation}
where $\delta^{M+N}(x^{\alpha}-x^{\alpha}(\sigma^{i}))$ is the 
Dirac delta function defined on the entire spacetime 
$({\cal M},g_{ab})$ by 
\begin{equation}
  \int_{\cal M}d^{N+M}x \; \delta^{M+N}(x^{\alpha}-x^{\alpha}(\sigma^{i}))
  = 1.
\end{equation}
Choosing the coordinate system $\{\xi^{\alpha},\sigma^{i}\}$, we
obtain
\begin{equation}
  \label{eq:NG-energy-momentum-tensor-in-xi-sigma}
  T^{ab}\sqrt{-g} =
  \mu \delta^{N}(\xi^{\alpha}) 
  \sqrt{-q} q^{ab} (\sigma_{i},\xi^{\alpha}=0).
\end{equation}

%*********************************************\\

Here, we note that the motion of the membrane is characterized by
$N$ functions $\xi^{\alpha}$, because the world sheet of the
membrane is the surface on which all $\xi^{\alpha}$ are constant.
If we consider the perturbed world sheet, we need only consider
the perturbations of the functions $\xi^{\alpha}$ and the
spacetime metric $g_{ab}$, as in the main text. 
From these perturbations, we can directly calculate the
perturbation of the intrinsic metric $q_{ab}$ of the membrane
world sheet.

%*********************************************\\

%%%%%%%%%%%%%%%%%%%%%%%%%%%%%%%%%%%%%%%%%%%%%%%%%%%%%%%%%%%%%%%%%%%%%
\subsection{Regularized Nambu-Goto membrane}
\label{sec:Appendix-thick-3}

%*********************************************\\

On the basis of the above energy-momentum tensor,
(\ref{eq:NG-energy-momentum-tensor-in-xi-sigma}), for an
infinitesimally thin Nambu-Goto membrane, we consider the
energy-momentum tensor for a regularized Nambu-Goto membrane with
which the dynamics of an infinitesimally thin Nambu-Goto membrane are
unchanged.
The energy-momentum tensor
(\ref{eq:NG-energy-momentum-tensor-in-xi-sigma}) has the following
properties:
(i) $T^{ab}$ is proportional to the induced metric $q^{ab}$ on
the world sheet of the membrane; and 
(ii) the support of $T^{ab}$ is confined to the world sheet (which
is proportional to $\delta^{N}(\xi^{\alpha})$). 
These properties should be separated when we consider a regularized
membrane.
We note that the first property is essential to the equation of
motion derived from the divergence of the energy-momentum tensor.
The second property simply reflects the fact that the membrane is
infinitesimally thin. 
Because we will study the dynamics of Nambu-Goto membrane using
a regularized membrane, we need only change property (ii) leaving
property (i) as it is.

%*********************************************\\

Now, we consider the energy-momentum tensor for the regularized
membrane with which the dynamics of this membrane are the same as
those of the Nambu-Goto membrane.
This tensor is the following:
\begin{equation}
  \label{eq:replaced-NG-Tmunu}
  T^{ab} = - \rho q^{ab}.
\end{equation}
Here, $\rho$ is a scalar function on ${\cal M}$ whose support is
the compact region ${\cal D}$ in the complement space of $\Sigma_{M}$.
The ``thick string world sheet'' is given by ${\cal D}\times\Sigma_{M}$.
Physically, the function $\rho$ represents the energy density of the
Nambu-Goto membrane. 
The usual line energy density $\mu$ in (\ref{eq:NG-action-NG-mem-1})
is given by 
\begin{equation}
  \mu = \int_{{\cal D}} d^{N}\xi \sqrt{\gamma} \rho.
\end{equation}
Similarly, according to the definition (\ref{eq:q-defE5}), $q^{ab}$ is
extended to the region ${\cal D}\times\Sigma_{M}$. 
The extended tensor $q^{ab}$ becomes a tensor of rank $M$ on 
${\cal D}\times\Sigma_{M}$. 
The extrinsic curvature $K_{abc}$ is also extended to a tensor on
${\cal D}\times\Sigma_{M}$ by substituting the extended $q_{ab}$ into
Eq.~(\ref{eq:K_abc-def}).
These extended versions of $q^{ab}$ and $K_{abc}$ are regarded as the
induced metric and the extrinsic curvature of each hypersurface
$\xi^{\alpha}=const.$ at each point on ${\cal D}$, respectively.  
As shown in the main text (Eqs.~(\ref{eq:NG-continuity-eq}) and
(\ref{eq:NG-eq-of-motion})), the divergence of the energy 
momentum tensor (\ref{eq:replaced-NG-Tmunu}) guarantees that the
dynamics of the regularized membrane are the same as those of the
Nambu-Goto membranes, though the additional continuity equation should
be taken into account.

%*********************************************\\

Finally, we comment on the meaning of the ``thickness'' of the
regularized Nambu-Goto membrane.
This ``thickness'' is determined by the length that
characterizes the support ${\cal D}$ of the energy density $\rho$.
For example, the region ${\cal D}$ may be chosen as the closure of the
$N$-dimensional open ball of radius $r_{*}$ in an appropriate chart.
The natural choice of the measure of the radius $r_{*}$ is the
circumferential radius of this open ball. 
With this choice, the original world sheet of the infinitesimally
thin Nambu-Goto membrane is obtained by choosing $r_{*}=0$ and the 
``thickness'' of membrane is characterized by the radius $r_{*}$.
We emphasize that we consider the situation in which the
thickness $r_{*}$ is much smaller than the curvature scale of the
bending membrane in order to obtain the dynamics of a thin membrane
and also to avoid Israel's paradox.
This approach was applied to an infinite string in our previous
works\cite{kouchan-string-1,kouchan-string-2}.

%*********************************************\\

%%%%%%%%%%%%%%%%%%%%%%%%%%%%%%%%%%%%%%%%%%%%%%%%%%%%%%%%%%%%%%%%%%%%%
\section{Perturbative Ricci curvatures}
\label{sec:Curvature-Appendix}
%%%%%%%%%%%%%%%%%%%%%%%%%%%%%%%%%%%%%%%%%%%%%%%%%%%%%%%%%%%%%%%%%%%%%

%*********************************************\\

Here, we list the perturbative curvatures at each order of the
perturbative treatment with Minkowski background spacetime.
As seen in \S\ref{sec:Perturbed-Einstein-Equations}, the
metric perturbations at each order are decomposed into gauge
invariant parts and gauge variant parts as
(\ref{eq:epsilon-metric-decomp}),
(\ref{eq:lambda-metric-decomp}), and
(\ref{eq:epsilonlambda-metric-decomp}). 
These decompositions are useful to calculate the perturbative
curvature at each order.
Further, the perturbative curvature at each order is represented by
the tensors ${\cal H}_{ab}^{\;\;\;\;c}$, ${\cal L}_{ab}^{\;\;\;\;c}$
and ${\cal K}_{ab}^{\;\;\;\;c}$ defined by (\ref{eq:calHabc-def}),
(\ref{eq:calLabc-def}) and (\ref{eq:calKabc-def}), respectively, 
and gauge variant part ${}^{(\epsilon)}X_{a}$ and
${}^{(\lambda)}X_{a}$ of the metric perturbation.

%*********************************************\\

The perturbative curvatures of each order are given by 
\begin{eqnarray}
  \left.
    \frac{\partial}{\partial\epsilon}R_{abc}^{\;\;\;\;\;\;d} 
  \right|_{\epsilon=\lambda=0}
  &=&
  - 2 \nabla_{[a} {\cal H}_{b]c}^{\;\;\;\;\;d}, \\
  \left.
    \frac{\partial}{\partial\lambda}R_{abc}^{\;\;\;\;\;\;d} 
  \right|_{\epsilon=\lambda=0}
  &=&
  - 2 \nabla_{[a} {\cal L}_{b]c}^{\;\;\;\;\;d}, \\
  \left.
    \frac{\partial}{\partial\epsilon} R_{ab}
  \right|_{\epsilon=\lambda=0}
  &=&
  - 2 \nabla_{[a}{\cal H}_{c]b}^{\;\;\;\;c}, \\
  \left.
    \frac{\partial}{\partial\lambda} R_{ab}
  \right|_{\epsilon=\lambda=0}
  &=&
  - 2 \nabla_{[a}{\cal L}_{c]b}^{\;\;\;\;c}, \\
  \left.
    \frac{\partial^{2}}{\partial\lambda\partial\epsilon} R_{ab}
  \right|_{\epsilon=\lambda=0}
  &=&
  - 2 \nabla_{[a} {\cal K}_{c]b}^{\;\;\;\;\;c}  
  + 2 {\cal H}_{[b}^{\;\;\;cd} {\cal L}_{c]ad} 
  + 2 {\cal L}_{[b}^{\;\;\;cd} {\cal H}_{c]ad}
  \nonumber\\
  && \quad
  + {\pounds}_{{}^{(\epsilon)}X} \left(\left.\frac{\partial}{\partial\lambda}
    R_{ab}\right|_{\epsilon=\lambda=0} \right)
  - {\cal H}^{cd} \left(\left. 
      \frac{\partial}{\partial\lambda} R_{acbd} \right|_{\epsilon=\lambda=0}
  \right)
  \nonumber
  \\
  && \quad
  + {\pounds}_{{}^{(\lambda)}X} \left(\left.\frac{\partial}{\partial\epsilon}
    R_{ab}\right|_{\epsilon=\lambda=0} \right)
  - {\cal L}^{cd}
  \left(
  \left.\frac{\partial}{\partial\epsilon}
    R_{acbd}\right|_{\epsilon=\lambda=0}
  \right). 
  \label{eq:ep-lam-order-Ricci-curvature-without-gauge-fix}
\end{eqnarray}
We can confirm these formulae in the straightforward way.
Using these formulae, we evaluated the Einstein equations order
by order in the main text.

%%%%%%%%%%%%%%%%%%%%%%%%%%%%%%%%%%%%%%%%%%%%%%%%%%%%%%%%%%%%%%%%%%%%%
\section{Harmonics for the mode expansion}
\label{sec:harmonics-Appendix}
%%%%%%%%%%%%%%%%%%%%%%%%%%%%%%%%%%%%%%%%%%%%%%%%%%%%%%%%%%%%%%%%%%%%%

To carry out the evaluation of the perturbation at each order,
it is useful to consider the mode expansion of the perturbative
variables. 
In this appendix, we introduce the harmonics for the expansion.
The harmonics introduced here are useful  to analyze the
perturbative oscillations of an infinite string, as done in the main
text.

%*********************************************\\

In the main text, we assumed that a string without oscillations is
straight and that the intrinsic metric on the string 
world sheet is given by
\begin{equation}
  q_{ab} = - (dt)_{a}(dt)_{b} + (dz)_{a}(dz)_{b}.
\end{equation}
This metric is also that on ${\cal M}_{1}$ in the main text.
We consider the mode expansion associated with the symmetries of this 
metric $q_{ab}$.
The natural scalar harmonic function on the string world sheet is
given by
\begin{equation}
  S := e^{-i\omega t + i k_{z} z}.
\end{equation}
From this scalar harmonic function, we can define the vector and
tensor harmonics.

%*********************************************\\

To consider the vector and tensor harmonics, it is necessary to
distinguish the perturbative modes according to whether or not
$\kappa$, defined by
\begin{equation}
  \label{eq:kappa-def}
  \kappa^{2} := \omega^{2} - k_{z}^{2},
\end{equation}
vanishes.
The $\kappa=0$ modes correspond to the perturbative modes
propagates along the string, and the $\kappa\neq 0$ modes correspond
to the other dynamical modes.
We should also comment that a different treatment is necessary
to consider cylindrical and stationary perturbations, as seen in
\S\ref{sec:Perturbed-Einstein-Equations}.

%*********************************************\\

%%%%%%%%%%%%%%%%%%%%%%%%%%%%%%%%%%%%%%%%%%%%%%%%%%%%%%%%%%%%%%%%%%%%%
\subsection{$\kappa\neq 0$ mode harmonics}
\label{sec:kappaneq0-mode-harmonics}

First, we introduce the $\kappa\neq 0$ mode harmonics.
Here, we denote all perturbative variables formally by $Q$.
Let us consider the perturbative variables $Q$ which satisfy the
equations
\begin{equation}
  \label{eq:kneq0-condition}
  {\cal D}_{c} Q \neq 0, \quad  {\cal D}^{c}{\cal D}_{c} Q \neq 0.
\end{equation}
The first condition in Eqs.~(\ref{eq:kneq0-condition}) implies that $Q$
is not constant and the second condition implies that the variable $Q$
is expanded by the eigen functions with the non-vanishing eigen value
$\kappa$ [see Eq.(\ref{eq:kappa-def})] of the derivative operator
${\cal D}^{c}{\cal D}_{c}$. 
According to the conditions (\ref{eq:kneq0-condition}), the
$\kappa\neq 0$ mode perturbative variables are defined.
If the perturbative variable $Q$ is a scalar, vector or tensors of
second rank on ${\cal M}_{1}$ , and if it satisfies
Eqs.~(\ref{eq:kneq0-condition}), it can be expanded in the harmonics 
\begin{eqnarray}
  \label{S-def-kneq0}
  S &:=& e^{-i\omega t + i k_{z}z}, \\
  \label{Vo1-def-kneq0}
  V_{(o1)}^{a} &:=& \epsilon^{ab}{\cal D}_{b}S, \\
  \label{Ve1-def-kneq0}
  V_{(e1)}^{a} &:=& q^{ab}{\cal D}_{b}S, \\
  \label{Te0-def-kneq0}
  T_{(e0)ab} &:=& \frac{1}{2} q_{ab} S, \\
  \label{Te2-def-kneq0}
  T_{(e2)ab} &:=& \left({\cal D}_{a} {\cal D}_{b} 
    - \frac{1}{2} q_{ab} {\cal D}^{c} {\cal D}_{c}\right)S, \\
  \label{To2-def-kneq0}
  T_{(o2)ab} &:=& - \epsilon_{c(a}{\cal D}_{b)}{\cal D}^{c} S,
\end{eqnarray}
where $\epsilon^{ab} = q^{ac}q^{bd}\epsilon_{cd}$ and
$\epsilon_{ab} = (dt)_{a}(dz)_{b} - (dz)_{b}(dt)_{a}$ are
two-dimensional totally antisymmetric tensors.

%%%%%%%%%%%%%%%%%%%%%%%%%%%%%%%%%%%%%%%%%%%%%%%%%%%%%%%%%%%%%%%%%%%%%
\subsection{$\kappa=0$ mode harmonics}
\label{sec:kappa=0-mode-harmonics}

Next, we introduce the $\kappa=0$ mode harmonics.
Let us consider the perturbative variables, which are denoted
formally by $Q$, satisfy the conditions  
\begin{eqnarray}
  \label{eq:k=0-condition}
  {\cal D}_{c} Q \neq 0, \quad  {\cal D}^{a}{\cal D}_{a} Q = 0.
\end{eqnarray}
The first condition in Eqs.~(\ref{eq:k=0-condition}) implies that $Q$
is not constant and the second condition implies that the variable $Q$
is expanded by the eigen functions with the vanishing eigen value 
$\kappa$ [see Eq.(\ref{eq:kappa-def})] of the derivative operator the
derivative operator ${\cal D}^{c}{\cal D}_{c}$.
According to the conditions (\ref{eq:k=0-condition}), the $\kappa=0$
mode perturbative variables are defined.
The scalar harmonic function for the $\kappa=0$ mode perturbations is
given by Eq.~(\ref{S-def-kneq0}) with the condition $\kappa=0$. 
To introduce vector and tensor harmonics of $\kappa=0$ mode, we
first introduce null vectors $k^{a}(=q_{b}^{\;\;a}k^{b})$ and
$l^{a}(=q_{b}^{\;\;a}l^{b})$ defined by 
\begin{equation}
  k_{a} := - i {\cal D}_{a}S, \quad k^{a}k_{a} = 0, \quad
  l_{a}k^{a} := - 2 \omega^{2}, \quad l^{a}l_{a} = 0.
\end{equation}
Using these null vectors, we introduce vector and tensor harmonics as
follows:
\begin{eqnarray}
  \label{eq:k0-scalar-harmonics}
  S &:=& e^{-i\omega (t + \nu z)}, \\
  \label{eq:k0-even-vector-harmonics}
  V_{(e1)}^{a} &:=& q^{ab} {\cal D}_{b}S, \\
  \label{eq:k0-l-mode-vector-harmonics}
  V_{(l1)}^{a} &:=& i l^{a} S, \\ 
  \label{eq:k0-trace-tensor-harmonics}
  T_{(e0)ab} &:=& \frac{1}{2} q_{ab} S, \\
  \label{eq:k0-even-tensor-harmonics}
  T_{(e2)ab} &:=& {\cal D}_{a}{\cal D}_{b} S, \\
  \label{eq:k0-l-mode-tensor-harmonics}
  T_{(l2)ab} &:=& - l_{a}l_{b} S, 
\end{eqnarray}
where $\nu=\pm 1$.
If the perturbative variable $Q$ is a scalar, vector or tensor
of second rank on ${\cal M}_{1}$, and if $Q$ satisfies the conditions 
(\ref{eq:k=0-condition}), it can be expanded in the harmonics
(\ref{eq:k0-scalar-harmonics})--(\ref{eq:k0-l-mode-tensor-harmonics}).

%%%%%%%%%%%%%%%%%%%%%%%%%%%%%%%%%%%%%%%%%%%%%%%%%%%%%%%%%%%%%%%%%%%%%
%%%%%%%%%%%%%%%%%%%%%%%%%%%%%%%%%%%%%%%%%%%%%%%%%%%%%%%%%%%%%%%%%%%%%

\end{document}